\documentclass[trackchanges]{aastex701}
\usepackage{dirtree}
\usepackage{multirow}
\usepackage{makecell}
\usepackage{booktabs}
\usepackage{comment}

\newcommand{\Mch}{M_{\mathrm{Ch}}}

\begin{document}

\title{First-Principles Turbulence-Driven Deflagration-to-Detonation Transition Mechanism 
for Near-Chandrasekhar Mass White Dwarf Progenitors}

\author[orcid=0009-0001-6426-1707,sname='Patel']{Krut Patel}
\affiliation{University of Massachusetts Dartmouth, Physics Department}
\email{kpatel29@umassd.edu}  

\author[orcid=0000-0002-1616-2245,sname='Dongre']{Akshay Dongre} 
\affiliation{University of Massachusetts Dartmouth, Physics Department}
\email{adongre1@umassd.edu}

\author[orcid=0000-0001-8077-7255,sname='Fisher']{Robert Fisher}
\affiliation{University of Massachusetts Dartmouth, Physics Department}
\email[show]{robert.fisher@umassd.edu}

\author[orcid=0000-0002-4605-7948,sname='Poludnenko']{Alexei Poludnenko}
\affiliation{University of Connecticut, School of Mechanical, Aerospace, and Manufacturing Engineering}
\email{alexei.poludnenko@uconn.edu}

\author[orcid=,sname='Gamezo']{Vadim Gamezo}
\affiliation{Naval Research Laboratory, Laboratories for Computational Physics \& Fluid Dynamics}
\email{vadim.gamezo@nrl.navy.mil}

\author[orcid=0000-0003-2021-8536,sname='Ugalino']{Mark Ugalino}
\affiliation{University of Maryland, Physical Sciences Complex, Stadium Drive}
\email{ugalino@umd.edu}

\author[orcid=0000-0002-0885-8090,sname='Byrohl']{Chris Byrohl}
\affiliation{Universität Heidelberg, Zentrum für Astronomie, Institut für Theoretische Astrophysik}
\email{chris.byrohl@uni-heidelberg.de}


\begin{abstract}

Type Ia supernovae (SNe Ia) play an important role throughout astrophysics, most notably as standardizable cosmological candles. Yet, their stellar progenitors and explosion mechanism remain areas of active investigation. For decades, the canonical model for normal brightness SNe Ia used in cosmology was a carbon-oxygen white dwarf (WD) accreting from a non-degenerate stellar companion, approaching the Chandrasekhar mass ($M_{\rm Ch})$. Previously, all models of near-$M_{\rm Ch}$ SNe Ia invoked an {\it ad hoc} assumption on the critical process of detonation initiation, and could therefore be tuned to a variety of outcomes. Here, we present global 3D hydrodynamical simulations of near-$M_{\rm Ch}$ progenitors, which incorporate, for the first time, a laboratory-validated {\it ab initio} mechanism for the turbulence-driven deflagration-to-detonation transition (tDDT). 
The tDDT detonation mechanism is highly efficient, leading to detonation initiation which is prompt in comparison to most prior work. 
Despite spanning a factor of six in central ignition density and qualitatively distinct ignition topologies, all models converge on nearly identical synthetic spectra at peak luminosity, spectroscopically matched to the overluminous SN 1999aa. The turbulence-driven Chapman-Jouguet criterion drives each progenitor to a common detonation configuration from diverse initial conditions, providing a physical foundation for the ignition-insensitive detonation outcomes implicit in the empirical standardizability of SNe~Ia.
This provides the first physically motivated, self-consistent pathway for delayed detonation in SNe Ia simulations. Further work is necessary to understand how this mechanism might  produce more delayed detonation initiation and potentially fail, thereby yielding SNe Iax.

\end{abstract}



\section{Introduction}
\label{chap:intro}

Type~Ia supernovae (SNe Ia) are luminous thermonuclear explosions of carbon--oxygen white dwarfs (WD) in binary systems~\citep{hoylefowler60}. 
Observations of high-redshift SNe~Ia led directly to the discovery of the accelerating Universe~\citep{Riess_1998,schmidt_1998,perlmutter_1999}. Beyond cosmology, SNe Ia are the primary contributors of iron-group elements to galaxies, ejecting Fe, Ni, and Mn into the interstellar medium~\citep{kobayashietal06}. Their optical spectra, characterized by persistent Si~II absorption and strong iron-group features, directly indicate nuclear burning in both the nuclear statistical equilibrium (NSE) and quasi-equilibrium (NSQE) regimes ~\citep{2021ApJ...922..205A, seitenzahltownsley17}.  

Equally, SNe~Ia serve as natural testbeds for 
high-energy astrophysics and turbulent combustion. The fundamental physics of their turbulent nuclear combustion connects with the astrophysical observables through their light curves and spectra~\citep{van_Rossum_2016}. 
Despite the central role of SNe Ia in modern astrophysics, the questions of their progenitor channels and explosion mechanisms remain largely open~\citep{maoz2014observational}. Two broad channels are recognized: in the single-degenerate channel, a WD accretes from a non-degenerate companion until reaching the Chandrasekhar mass, igniting carbon in the core~\citep{whelan1973binaries,nomoto1982accreting}; in the double-degenerate channel, two WDs merge following orbital decay~\citep{iben1984supernovae,webbink84}. Abundance ratios (in particular, [Cr/Fe]) inferred from the galactic supernova remnants are sensitive to the progenitor central density and metallicity, and point towards a Chandrasekhar-mass progenitor in the SN Ia SNR 3C 397~\citep{yamaguchi2015chandrasekhar,daveetal17}. On the other hand, the existence of Gaia hypervelocity WDs \citep {shen2018three} strongly suggests that some SNe Ia originate from sub-$\Mch$ WD progenitors. Thus it seems increasingly likely that multiple progenitor channels contribute to the observed population of SNe Ia.

While SNe Ia have been remarkably successful as standardizable candles \citep{riess..1998AJ....116.1009R, perlmutter_1999, Brout2022}, next-generation surveys (LSST, Roman) will reach precision where currently subdominant systematics, including channel-dependent intrinsic luminosity variations tied to progenitor and explosion physics, may bias the dark energy inference if the channel mix evolves with redshift \citep{Maoz2014, Rigault2020, Jones2018}. Resolving the explosion mechanism in full star simulations is therefore increasingly tied to the precision-cosmology program, and is the focus of this paper.


This work focuses on the explosion physics within the near-$\Mch$ single-degenerate paradigm. In a near-$\Mch$ WD, carbon ignition produces a subsonic deflagration~\citep{nomoto1984accreting}. How this flame propagates, and, crucially, whether it transitions to a supersonic detonation, largely determines the explosion outcome \citep{khokhlov91, khokhlovetal97, roepkeetal07}. Pure deflagration cannot reproduce normal SNe Ia, as they burn too slowly, expand the star excessively, and yield explosions that are too faint~\citep{gamezoetal04}. Instead, such explosions are largely consistent with the peculiar Iax subclass of SNe Ia, including the prediction that they should leave behind gravitationally bound, kicked remnants similar to LP 40-365 ~\citep{ropke2012constraining, jordanetal12b, raddietal18}. 

\citet{khokhlov91} originally proposed the delayed-detonation model that resolves the shortcomings of a pure deflagration with a two-stage process: a subsonic deflagration stage that pre-expands the star, followed by a supersonic detonation that incinerates the remaining fuel. The delayed-detonation model reproduces key observational features, including luminosity and stratified ejecta. While substantial work has been devoted to exploring the physics of the transition to a detonation  \citep {niemeyerwoosley97, woosley07}, the physical mechanism has remained unsolved until recently. In the absence of a first-principles mechanism, previous large-scale simulations have imposed detonation by prescribing a critical density threshold at the flame front, typically $\rho_{\mathrm{DDT}} \approx 10^{7} \, \text{g/cm}^{3}$~\citep{gamezoetal05,seitenzahletal13}. The tunable transition density approach introduces significant uncertainty, and makes it challenging to infer progenitor properties from the observed light curves and spectra.

Recent advances in combustion theory have demonstrated a new mechanism for the deflagration-to-detonation transition (DDT). Terrestrial experiments and direct numerical simulations of flames (both chemical and thermonuclear)~\citep{poludnenkoetal11,poludnenkoetal19} showed that sufficiently strong turbulence can drive subsonic flames to self-accelerate to a detonation. One dimensional conservation laws prohibit a steady deflagration above a critical threshold: the Chapman-Jouguet deflagration speed $S_{\mathrm{CJ}}$. However, turbulence wrinkles the flame surface and accelerates the burning and can cause a flame to exceed this critical Chapman-Jouguet deflagration speed and initiate supersonic burning and a detonation.
In the near-$\Mch$ SNe Ia, Rayleigh--Taylor (RT) instabilities of the subsonically propagating flame front naturally generate turbulence~\citep{zingale2011convective}. The wrinkling of the flame becomes more enhanced as it burns buoyantly outward through the star and the flame speed drops,  making this turbulence-driven DDT (tDDT) mechanism nearly inevitable in any purely hydrodynamic evolution of the flame surface.

Here, we present the first full-star, three-dimensional hydrodynamic simulations of the near-$\Mch$ progenitors, in which the DDT is triggered by the newly-validated tDDT mechanism, rather than imposed by prescription. We describe the precise hydrodynamic conditions, which lead to a detonation, as well as compute detailed nucleosynthetic yields and synthetic spectra. The synthetic spectra enable direct comparison with the observations and assessment of the tDDT mechanism's viability for explaining SNe Ia.

Our initial stellar models, the choice of the initial ignition parameters, and the methodologies of the implementation of the tDDT criterion are summarized in Section~\ref{chap:method}. We discuss the flame morphologies for each model and the results of their nucleosynthesis in Section~\ref{chap:results}, and later compare the spectral features against observed SNe Ia events. Finally, in Section~\ref{chap:conclusion}, we summarize our findings and discuss their implications.

\section{Methodology}
\label{chap:method}

\subsection{Initial WD models, hydrodynamics, flame modeling}
\label{subsec:init-models}
In the single-degenerate near-$\Mch$ scenario, a wide range of binary stars can lead to runaway carbon burning in the core of the primary WD. Prior work on binary population synthesis models has demonstrated that this range of the initial conditions of the binaries can be absorbed into a single one-dimensional parameter variation of the central density of the primary WD \citep {lesaffreetal06}. Our simulations begin at the onset of the deflagration runaway at the WD core.
\citet{lesaffreetal06} demonstrate ignition conditions from $2 - 5 \times 10^{9} \, \mathrm{g\,/\,cm^{3}}$, although the low end of their distribution appears to be artificially truncated. Other more recent work on X-ray observations of SNR 3C 397, which requires significant neutronization to produce the observed elemental abundances, suggests even higher densities may be required to produce the observed abundance ratios of Cr/Fe \citep {daveetal17, oshiroetal21}.  Thus, to fully explore the entire range of SNe Ia progenitor scenarios, we consider three near-\( M_\mathrm{Ch} \) WDs with low (LCD), standard (SCD), and high (HCD) central densities: 
$1 \times 10^{9} \, \mathrm{g\,/\,cm^{3}}$, $2.2 \times 10^{9} \, \mathrm{g\,/\,cm^{3}}$ and $6.0 \times 10^{9} \, \mathrm{g\,/\,cm^{3}}$, respectively.  The standard density values represent the canonical ignition densities adopted in many delayed-detonation studies~\citep{ropkeetal07, seitenzahletal13b}, while the low and high density models probe the extremes of possible single-degenerate progenitors.

Initial 1D WD profiles were generated with an adiabatic temperature gradient representing the inner convective core during the simmering phase leading up to the runaway, with a maximum temperature $T_{\mathrm{c}} = 7.0 \times 10^{8}$ K, and a cold base temperature \(T_\mathrm{base}\sim 1.0 \times 10^7\) K. The composition of the WD is chosen to be homogeneous 50/50 C/O by mass for all models. 1D white dwarf profiles were first mapped onto a 3D Cartesian grid 
using mass-weighted averages over spherical shells of width $\Delta r$. 
The base stellar models are 
nearly in hydrostatic equilibrium, with a fractional virial error \(\sim 1 \times 10^{-5}\). 

The ignition phase of the deflagration remains poorly understood. The turbulent convective flow during the century-long simmering phase is computationally prohibitive to capture accurately. The handful of simulations of the thermonuclear runaway generally demonstrate the development of a single radially offset ignition bubble ~\citep{zingaleetal09, nonakaetal12}. However, prior simulations have also considered multiple ignition bubbles \citep {seitenzahletal13b}.  We therefore treat the initial ignition configuration, including number of ignition kernels, their position, and radius, as adjustable parameters in our models (Table 1). This approach allows us to explore the sensitivity of tDDT outcomes to ignition conditions.

\begin{deluxetable*}{lcccc}
\tablewidth{0pt}
\tablecaption{Cases presented with their ignition parameters. For the multipoint ignition case N100r16, the offset ranges from 0.3 km to 136 km, decreasing flame bubble number density with increasing offset. \label{table:runs}}
\tablehead{
\colhead{Run ID} & \colhead{$\rho_c$} & \colhead{Ignition Offset} & \colhead{Ignition Radius} & \colhead{Ignition points} \\
\colhead{} & \colhead{(g cm$^{-3}$)} & \colhead{(km)} & \colhead{(km)} & \colhead{}
}
\startdata
LOW-O12R32  & $1.0 \times 10^9$ & 12 & 32 & 1 \\ 
STD-O12R32  & $2.2 \times 10^9$ & 12 & 32 & 1 \\ 
STD-O100R32 & $2.2 \times 10^9$ & 100 & 32 & 1 \\ 
STD-N100R16 & $2.2 \times 10^9$ & $\sim 0.3$ to $\sim 136$ & 16 & 100 \\
HIGH-O12R32 & $6.0 \times 10^9$ & 12 & 32 & 1 \\ 
HIGH-O80R32 & $6.0 \times 10^9$ & 80 & 32 & 1 
\enddata
\end{deluxetable*}

We use \texttt{FLASH} 4.3 \citep{fryxell2000flash, dubeyetal09, Dubey14} to perform  3D hydrodynamical simulations. 
\texttt {FLASH} is an Eulerian, adaptive-mesh, parallel, multiphysics simulation framework. 
Hydrodynamics is evolved using the unsplit higher-order Godunov solver~\citep{calderetal07}.
Self-gravity is treated with an improved multipole Poisson solver~\citep{couch_2013}, with the origin  fixed at the mean square-density-weighted position, and the gravitational potential centered at cell faces.  Diode boundary conditions are imposed on the potential, retaining terms through $l = 6$.
\texttt{FLASH} employs the Helmholtz free energy equation of state, which accounts for the local thermodynamic equilibrium contributions of non-degenerate nuclei, and arbitrary degeneracy and relativity of electrons and positrons~\citep{timmeswoosley92}. The nuclear-burning flame is captured using a three-stage artificially-thickened flame model based on the advection-diffusion-reaction equation~\citep{calderetal07}. The flame module incorporates nuclear energy generation calibrated against the self-heating reaction networks, and also includes neutronization, electron screening, and Coulomb corrections~\citep{townsleyetal07,townsleyetal09,townsleyetal16}. Three separate scalar fields track the progression of carbon burning, burning to NSQE, and burning to NSE, allowing us to follow the evolution from a subsonic deflagration through potential transition to a supersonic detonation.

Adaptive mesh refinement strategies based on the cell density and specific nuclear energy generation rate allow us to resolve the ignition kernels and the flame at the maximum finest linear resolution of 4 km, thus capturing the Rayleigh-Taylor and Kelvin-Helmholtz instabilities, while maintaining the spherical geometry of the WD at a comparatively lower resolution of 16 km. The domain size was chosen to be \(2.62144 \times 10^{5}\,\mathrm{km}\) in each direction to follow the high-velocity explosion ejecta until free expansion. Furthermore, once the detonation is formed, the gravitationally unbound ejecta are evolved with a constant mass refinement scheme, which enables a gradient of resolution in the grid structure, with the highest resolution within regions with the highest mass. Eulerian simulations require non-vacuum conditions, thus around the WD, the fluff temperature and density were set to $T_{\mathrm{fluff}} = 1.0 \times 10^{6} \, \mathrm{K}$ and $\rho_{\mathrm{fluff}} = 1.0 \times 10^{-3} \,\mathrm{g/cm^{3}}$. The lowest grid resolution was set to 4096 km initially in the fluff region.

\subsection{tDDT mechanism}
\label{subsec:tddt-method}

The key insight underlying the tDDT mechanism~\citep{poludnenkoetal11, poludnenkoetal19} is that turbulent flames only need to exceed the Chapman-Jouguet (CJ) deflagration speed $S_\mathrm{CJ}$, not the detonation speed, to trigger a pressure runaway that may lead to DDT. This represents a critical departure from earlier models that assumed distributed burning and detonation-strength shocks were required for DDT~\citep{niemeyerwoosley97,khokhlovetal97}. To appreciate why this seemingly slower threshold is sufficient, it is instructive to consider the CJ condition itself.

Consider a combustion wave -- either a deflagration or a detonation -- propagating through a reactive medium. In the reference frame co-moving with the wave front, fresh fuel flows in from upstream and combustion products flow out downstream. The CJ condition is satisfied when the downstream (product-side) flow velocity in that frame equals local sound speed. This sonic-point condition is the key ingredient of the CJ physics: when the downstream flow is exactly sonic, rarefaction waves cannot propagate upstream through the products to weaken the wave.

The CJ condition can be satisfied at two distinct speeds, giving rise to two physically distinct CJ solutions. The first solution (CJ detonation) corresponds to a supersonic wave led by a strong shock that ignites the material. The CJ detonation solution defines the minimum possible velocity $D_\mathrm{CJ}$ of a steady-state detonation wave. The second solution (CJ deflagration) corresponds to a subsonic wave propagated by thermal conduction and diffusion (or turbulent transport) without shocks, and it defines the maximum possible velocity $S_\mathrm{CJ}$ of a steady-state deflagration wave. There are no steady-state solutions for wave velocities between $D_\mathrm{CJ}$ and $S_\mathrm{CJ}$. Both solutions satisfy conservation laws~\citep{landaulifshitz59} and include the energy released by chemical or thermonuclear reactions that drive both deflagrations and detonations. The energy release uniquely defines $D_\mathrm{CJ}$ and $S_\mathrm{CJ}$, but neither $D_\mathrm{CJ}$, nor $S_\mathrm{CJ}$ depend on the rate of energy release. 

Waves propagating faster than $D_\mathrm{CJ}$ (overdriven detonations) are possible, but they cannot be sustained without external forcing, such as a piston, and will relax to the CJ detonation state. Thus, one-dimensional (1D) self-sustained steady-state detonations always propagate with the velocity $D_\mathrm{CJ}$. 

The propagation velocity $S$ of the deflagration waves (flames) can vary depending on the material properties, presence of turbulence, and the rate of energy release. In particular, steady-state turbulent flames can propagate at any velocities $S_T < S_\mathrm{CJ}$, but since there are no steady-state solutions for $S_T > S_\mathrm{CJ}$, faster flames become unstable and generate pressure waves. This can result in a pressure runaway that may eventually cause the combustion wave to transition to the faster steady-state solution, which is the CJ detonation. 

It can be shown that flames exceeding the CJ deflagration speed must burn on a timescale shorter than the sound-crossing timescale, i.e.,
\begin{equation}
\label{eq:tddt_criterion}
S_T > S_\mathrm{CJ} \quad \leftrightarrow \quad \dot{e} \gtrsim \frac{e}{t_s},
\end{equation}
where $\dot{e}$ is the energy release rate per unit volume, $e$ is the internal energy density, and $t_s = \delta_T/c_s$ is the sound-crossing time of the turbulent flame with thickness $\delta_T$~\citep{poludnenkoetal11}.

Equation~(\ref{eq:tddt_criterion}) represents the underlying physics: when $S_T > S_\mathrm{CJ}$, burning releases within one sound-crossing time an amount of energy comparable to (or greater than) the internal energy stored in the flame volume. This rapid energy deposition creates a pressure build-up that cannot be relieved acoustically, ultimately driving the formation of strong shocks. These shocks can then amplify through interactions with the turbulent flame until their strength reaches that of a CJ detonation and thus becomes sufficient to directly support detonation propagation~\citep{poludnenkoetal19}.

For tDDT to occur, turbulence must fold the flame and create sufficient amount of the flame surface for 
the resulting turbulent flame speed $S_T$ to exceed the CJ deflagration speed $S_\mathrm{CJ}$. Once $S_T > S_\mathrm{CJ}$, the flame consumes
fuel faster than upstream turbulence can resupply it, decoupling from the
ambient flow and driving a pressure build-up on the unburned side. This
transient, super-CJ burning phase is what ultimately launches the shocks that can ignite a detonation.

Two quantities control whether this condition is reached: the turbulent intensity $U_\mathrm{CJ}$ required to increase $S_T$ to $S_\mathrm{CJ}$, and the spatial scale (or flame brush thickness) $L_\mathrm{CJ}$, at
which that intensity must be sustained. In the flamelet regime, the turbulent flame speed at scale $L$ can be expressed in terms of the folded flame surface area $A_T$ as~\citep{poludnenkoetal11}
\begin{equation}
\label{eq:st_flamelet}
S_T = I_M S_L \frac{A_T}{L^2} = I_M S_L \frac{L}{\lambda_f},
\end{equation}
where $I_M \sim 1$ is the Markstein number, $S_L$ is the laminar flame speed, and $\lambda_f$ is the flame-polishing scale, i.e., the smallest scale, on which turbulence wrinkles the flame while burning smooths the flame out on scales below $\lambda_f$.


Combining Equation~(\ref{eq:st_flamelet}) with the velocity scaling in the quasi-steady Kolmogorov-type turbulence $U \propto L^{1/3}$ and the flame-polishing condition $U(\lambda_f) = \alpha I_M S_L$ yields the scale, at which $S_T = S_{CJ} = c_s/\alpha$
\begin{equation}
\label{eq:lcj_general}
L_\mathrm{CJ} = (\alpha I_M S_L)^2\, c_s\, \frac{l}{U_l^3},
\end{equation}
and the turbulent intensity at that scale
\begin{equation}
\label{eq:ucj}
U_\mathrm{CJ} = (\alpha I_M S_L)^{2/3}\, c_s^{1/3},
\end{equation}
where $c_s$ is the sound speed in hot products, $I_M \sim 1$ is the Markstein number, and $\alpha = \rho_f/\rho_p$ is the fuel-to-products density ratio (see \citet{poludnenkoetal19} for further details). Equations~(\ref{eq:lcj_general}) and~(\ref{eq:ucj}) are the operational criteria evaluated in our simulations: conditions for the onset of tDDT are considered to be satisfied in a local region when the local turbulent intensity exceeds $U_\mathrm{CJ}$ over a region of size $L_\mathrm{CJ}$.

The CJ deflagration criteria described above only provide the necessary, but not a sufficient condition. In order to sustain the pressure runaway long enough to amplify a shock to a strength necessary for DDT, burn-out time $t_B$ of the fuel inside the turbulent flame must exceed the sound-crossing time $t_s$ of the corresponding flame volume. It is only during this time that the flame speed can remain much larger than the characteristic turbulent velocity $U$, i.e., $S_T \gtrsim S_{CJ} \gg U$, before the flame returns to an equilibrium with the surrounding turbulence. More specifically, it can be shown that this requirement reduces to $t_B/t_s = \alpha$ \citep{poludnenkoetal11}. Therein lies the central
difficulty for thermonuclear DDT: chemical flames have $\alpha \sim 3$--$10$,
giving several sound-crossing times for pressure to build and directly ignite
a detonation, whereas degenerate thermonuclear flames have only
$\alpha \sim 1.2$--$2.0$ \citep{timmeswoosley92}. Resulting shock strengths in the latter 
are close to those that can be achieved in a constant-volume explosion and are thus too
weak to directly detonate the fuel. Therefore, thermonuclear tDDT in SNe Ia proceeds in two
stages: an initial shock launched by a super-CJ deflagration, followed by the
amplification of that shock through the interaction with the surrounding turbulent flame and compression
of additional reactive material \citep{poludnenkoetal19}. This two-stage picture forms the physical basis for the tDDT criterion implemented in our simulations.

The derivation of $L_\mathrm{CJ}$ and $U_\mathrm{CJ}$ above assumes that turbulence in the reactive flow can be approximated to be homogeneous, isotropic, Kolmogorov-type. While exothermic reactions can modify turbulence structure within flames primarily through thermal expansion and temperature-dependent viscosity \citep{Hamlington2011,Hamlington2012}, these effects are minimal in thermonuclear flames due to very low density ratios across the flame ($\alpha \lesssim 2$) and extremely small Prandtl numbers ($Pr \sim 10^{-5}$) in degenerate plasmas. 
Detailed numerical simulations have confirmed that Kolmogorov spectra are preserved in highly-resolved RT flame-turbulence interactions~\citep{Zingale2005}, supporting this conclusion. 
We have further confirmed that the turbulent kinetic energy spectra over the resolved inertial range in our models also exhibit an approximate power-law scaling close to $E(k) \propto k^{-5/3}$. The similarity of the spectra at different times suggests that the turbulent cascade remains approximately self-similar during the deflagration phase, even as the flame and large-scale flow evolve. This Kolmogorov velocity scaling supports the derivation of $L_\mathrm{CJ}$ and $U_\mathrm{CJ}$, since the subgrid turbulent intensity is ultimately controlled by an inertial cascade that is well represented by the standard cascade picture on the resolved scales. 


To implement the tDDT mechanism in our simulations, we must extract the subgrid-scale turbulent intensity from the resolved computational scales. A simple metric like the root-mean-square velocity within a given region accounts for all kinetic energy -- both vortical turbulent power as well as bulk radial expansions and compressions.  We therefore employ a calibrated differential operator, OP$_2$~\citep{colin2000}, which measures solenoidal (rotational) velocity component while filtering the dilatational (compressive) motions. This operator is defined as
\begin{equation}
\label{eq:op2}
u' = \text{OP}_2(\vec{u}) = c_2^h (\Delta)^3 \nabla^2 (\nabla \times \vec{u}),
\end{equation}
where $u'$ is the estimated turbulent intensity, $c_2^h$ is a calibration constant depending on the numerical method and the order $h$ of the finite-difference stencil, $\Delta$ is the grid spacing, and $\vec{u}$ is the local velocity field. The curl operator $\nabla \times \vec{u}$ extracts the vorticity, filtering out the dilatational modes associated with acoustic waves and the mean flow. The Laplacian $\nabla^2$ acts as a high-pass spatial filter, emphasizing small-scale fluctuations. The factor $(\Delta)^3$ ensures dimensional consistency.

\citet{jacksonetal2010} implemented this operator in \texttt{FLASH} via a fourth-order central differencing scheme to compute first- and second-order derivatives, and calibrating it to verify that it accurately reproduces the resolved simulated turbulent power. In our calculations, the operator is evaluated over a stencil of width $4h\Delta$, which serves as the effective length scale $L$ for the measurement of $u'$ in the tDDT criterion. This approach allows us to monitor locally whether the condition for the pressure runaway, Equation~(\ref{eq:tddt_criterion}), is satisfied, even in the presence of complex flow structures involving shocks, expansion waves, and strong mean flows.


A computational cell is marked for detonation triggering if two conditions are simultaneously satisfied: (i) the flame progress variable indicates that the cell lies within the active flame front, thus excluding spurious triggers in fully unburned or fully burned regions; and (ii) $ L/ L_\mathrm{CJ} \geq 1$, indicating that local turbulent flame volume exceeds the critical scale required for the pressure runaway. Here, the length scale ratio $L/ L_\mathrm{CJ}$  is
\begin{equation}
\label{eq:l_ratio}
\frac{L}{L_\mathrm{CJ}} = \frac{(u')^3}{c_s (\alpha I_M S_L)^2} = \frac{(u')^3}{U_{CJ}^{3}},
\end{equation}
where the local values of $u'$ (from OP$_2$), $c_s$, $\alpha$, $I_M$, and $S_L$ are used. The $L \sim 4h\Delta$ is the OP$_2$ measurement scale. Cells satisfying both criteria transition from a deflagration to a detonation. 
To prevent multiple triggers in nearby cells, only one trigger is allowed per the AMR grid block.
This implementation 
provides a physics-based, self-consistent approach to modeling DDT in SNe Ia explosions that can be applied across widely varying WD compositions, densities, and turbulent conditions. 

\subsection{Nucleosynthesis and Synthetic Spectra}
\label{sec:nucleo}
The 3-stage flame burner, which we employ, suffices to determine the total nuclear energy released during burning, but does not provide the detailed nucleosynthetic yields required for the radiative transfer post-processing. Therefore, calculations include passive Lagrangian tracer particles, which record the thermodynamic state of the fluid and thus can be used to calculate the nucleosynthetic yields. All models included $10^5$ tracer particles distributed proportionally by mass: each computational block receives a number of particles proportional to the mass fraction contained in it relative to the total WD mass. Specifically, if a block contains mass \(M_\mathrm{block}\) and the total WD mass is \(M_\mathrm{WD}\), that block receives approximately \(N_\mathrm{block} = N_\mathrm{total} \times (M_\mathrm{block}/M_\mathrm{WD})\) tracer particles. Due to the integer rounding in this allocation algorithm, the actual final number of particles may differ slightly from the target value. This mass-weighted distribution ensures that the tracer ensemble accurately represents the mass distribution of the ejecta, with denser regions containing proportionally more particles. 

These particles passively advect with the flow and record the underlying density and temperature profiles from the Cartesian grid at each hydrodynamic timestep. Each tracer particle's mass is computed from its allocation, such that the sum of all particle masses equals the total ejecta mass, ensuring mass conservation in the nucleosynthesis post-processing.

To quantify the transition to free expansion, we define a dimensionless homology criterion $\Delta$ as the standard deviation over the particle data of the Hubble parameter $v_r/r$ normalized by its mean. Here, $v_\mathrm{r}$ is the radial velocity and $r$ is the radial position of a tracer particle. This parameter quantifies the spread of the inverse flow timescale in the particle data. Once a successful detonation was achieved, and the WD became gravitationally unbound, simulations were continued until $\Delta \sim 0.1\%$. This condition ensures that the ejecta velocity field is nearly homologous (constant $v_\mathrm{r} / r$) corresponding to free expansion, which is required to ensure accurate mapping of the ejecta to the homologous velocity grid used in radiative transfer. 

The tracer particle data, recorded at each timestep for all particles, were then post-processed to reconstruct the individual particle thermodynamic histories. For each particle, we extract its particle tag number, time, temperature trajectory, and density trajectory from ignition through free expansion. These trajectories capture the full burning history experienced by different fluid elements as they undergo deflagration, possible DDT, and subsequent expansion.

Nucleosynthesis calculations along these trajectories were performed using the \texttt{Torch} code~\citep{timmes_1999}, which implements a 489-isotope reaction network including $\alpha$-particle reactions, proton and neutron captures, $\beta$-decays, and electron captures. 
For each tracer particle trajectory, the initial composition was set to 50/50 $^{12}$C/$^{16}$O by mass, consistent with the hydrodynamic initial conditions.

The final nucleosynthetic yields for each tracer particle include mass fractions of all 489 isotopes, including the distribution of iron-peak elements (noteably, Ni, Fe, Mn, and Cr), intermediate-mass elements (Si, S, Ca, Ar), and unburned material (C, O). The total $^{56}$Ni mass, a primary determinant of the SNe Ia peak luminosity, is computed by summing contributions from all tracer particles. The 3D nucleosynthetic yields from the ensemble of tracer particles were mapped onto a 1D spherical velocity mesh to prepare for radiative transfer calculations. This dimensional reduction from the 3D to 1D spherical symmetry is justified by the highly spherical structure of the ejecta in our models.

Radiative transfer calculations were performed using \texttt{SuperNu}~\citep{wollaeger_2013, wollaeger_2014}, a time-dependent, radiative transfer code designed for SNe Ia and other rapidly expanding transients. \texttt{SuperNu} employs a hybrid transport scheme combining Implicit Monte Carlo (IMC) and Discrete Diffusion Monte Carlo (DDMC) methods. IMC follows individual photon packets stochastically through the ejecta, accounting for absorption, emission, and scattering processes, while stabilizing the timestep through effective scattering. DDMC accelerates transport in optically thick regions deep within the ejecta by replacing many individual scattering events with diffusion steps, significantly improving the computational efficiency without sacrificing accuracy. \texttt{SuperNu} also implements an opacity regrouping scheme~\citep{wollaeger_2014} that bins lines by frequency and strength while preserving total opacity and improving computational efficiency.

The \texttt{SuperNu} calculations discretize the 1D ejecta model onto 100 velocity zones spanning \(1000\,\mathrm{km\,s^{-1}} \lesssim v \lesssim 30{,}000\,\mathrm{km\,s^{-1}}\), with radiative transfer computed from \(t = 1\,\mathrm{day}\) to \(t = 60\,\mathrm{days}\) post-explosion using logarithmically spaced timesteps to resolve both the rapid rise to maximum light and the subsequent decline phase. A total of \(2^{25}\) photon packets were emitted over the simulation to ensure statistical convergence. Emergent spectra were tallied into 512 logarithmically spaced wavelength bins covering $10^3$--$10^{4.5}$~\AA\ ($\sim$1000~\AA\ to 3.16~$\mu$m), spanning the far-ultraviolet through the near-infrared range, over which SNe~Ia radiate appreciably.
The resulting \texttt{SuperNu} output comprises time-dependent bolometric light curves and high-resolution synthetic spectra at multiple epochs directly comparable to observations for quantitative model assessment. 


To assess the fidelity of our theoretical models against observations, synthetic spectra were classified using the SuperNova Identification (SNID) code \citep{blondin2011snid}. SNID provides a robust, quantitative method for supernova classification by comparing an input spectrum against the extensive library of high-quality spectral templates compiled from well-observed SNe of various types and subtypes. 
The core methodology of SNID is a modified cross-correlation algorithm \citep{tonry_1979}, which simultaneously determines the best-fitting supernova type/subtype, redshift ($z$), and spectral phase (epoch) relative to maximum light.

For each comparison between an input spectrum and a template, SNID computes the cross-correlation function in the logarithmic wavelength space after normalizing both spectra. The goodness-of-fit is quantified by the \texttt{rlap} parameter:
\begin{equation}
\label{eq:rlap}
\texttt{rlap} = r \times \mathrm{lap},
\end{equation}
where \(r\) is the correlation peak height, measuring the similarity of spectral features, and \(\mathrm{lap}\) is the fractional wavelength overlap between the input and template spectra. Typical good matches have \(\texttt{rlap} > 5\). Higher \texttt{rlap} values indicate better matches. For SNe Ia classification, values \(\texttt{rlap} \gtrsim 8\) generally indicate robust type identification. For our analysis, synthetic spectra from each model were extracted at epochs spanning -3 to +3 days relative to peak light. Each spectrum was classified using SNID with redshift fixed to the rest frame at \(z = 0\),  
over a 2000--10,000~\AA\ range spanning the near-UV through near-IR. For general transport, a wider spectral grid is retained to capture the full bolometric flux and to properly bound the UV line-blanketing and near-IR transport at the spectral edges.

In addition to the best-fitting template, SNID output also provides a ranked list of top matches, allowing us to assess whether our models resemble peculiar SNe Ia or normal SNe Ia. We report the top two best-matching observed events and subtypes for each model, providing a direct observational classification of our theoretical predictions.


\section{Results}
\label{chap:results}

\subsection{Hydrodynamical Outcomes}
\label{sec:ejecta}
All six models achieve a successful detonation, demonstrating that the tDDT mechanism operates robustly across the range of progenitor central densities and ignition configurations explored here. Despite this shared outcome, the hydrodynamical evolution prior to and at the time of detonation initiation varies significantly between models, governed primarily by the progenitor's initial central density and the ignition geometry.

\begin{figure}
    \centering
    \includegraphics[width=0.7\textwidth]{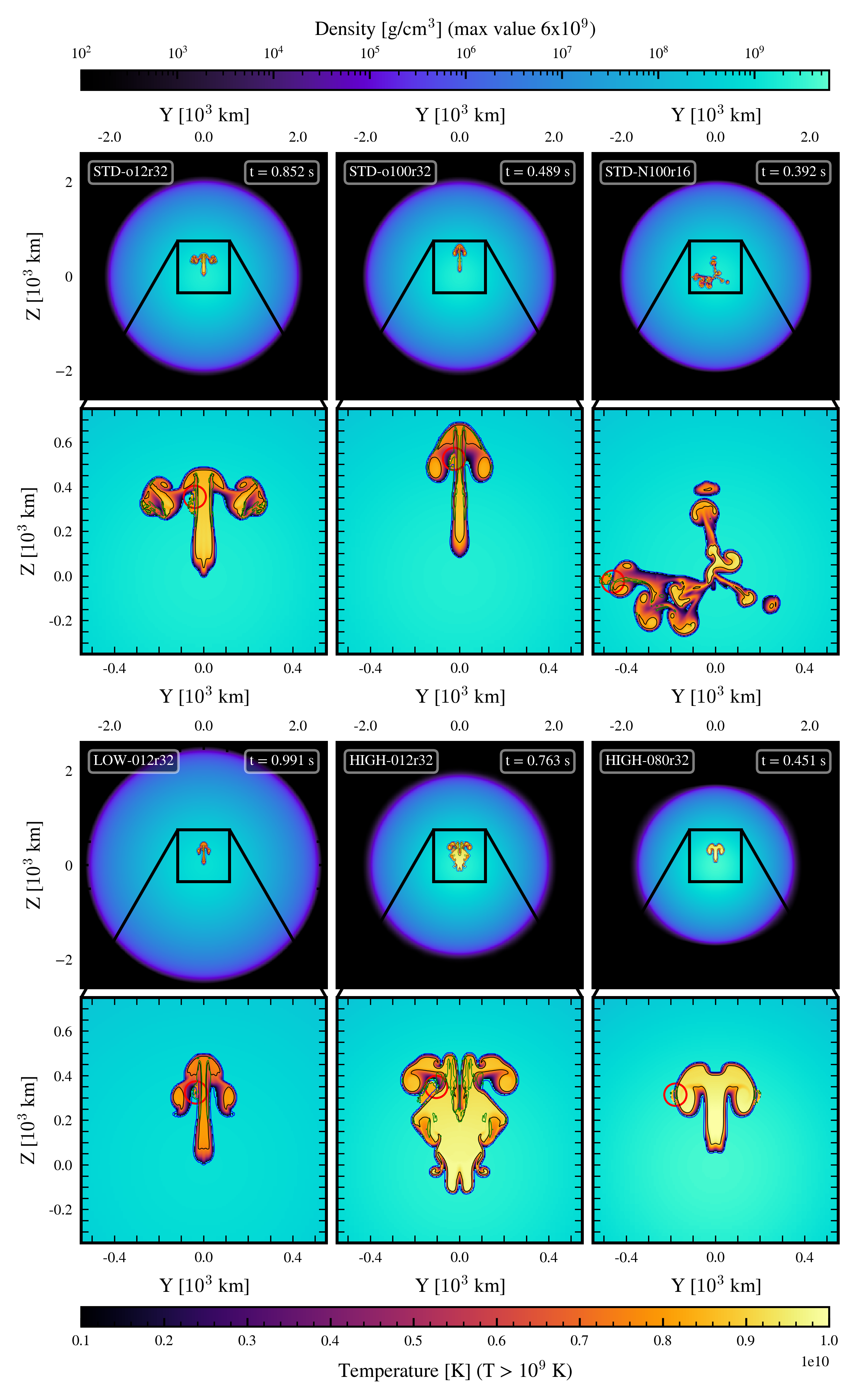}
    \caption{Slices in the $y$-$z$ plane for all six models at the onset of the first detonation trigger, shown at the times indicated in each panel. The upper subpanel of each pair shows the full WD on a density color scale (top colourbar; $\mathrm{g\,cm^{-3}}$; clipped at a maximum of 
    $6 \times 10^9\,\mathrm{g\,cm^{-3}}$), with an inset box indicating the region enlarged in the lower subpanel. The lower subpanel shows the zoomed flame structure on a temperature color scale (bottom colourbar; K; clipped at $T > 10^9\,\mathrm{K}$). In both subpanels, black contours mark the flame progress variable $= 1$ (fully burned material), blue contours mark the flame progress variable $= 0.001$ (leading edge of the flame front), and green contours mark $L/L_\mathrm{CJ} = 1$. Detonation is triggered where the green contours intersect the blue, indicating that the local turbulent flame volume exceeded the critical Chapman-Jouguet scale.}
    \label{fig:det_slice}
\end{figure}


All simulations are initiated with off-center ignition kernel(s) that seed subsonic flames. The burning front propagates radially outward, growing the flame bubble as freshly ignited material expands and rises buoyantly through the denser surrounding fuel. Once the bubble reaches a critical size, it becomes unstable to Rayleigh-Taylor (RT) and Kelvin-Helmholtz instabilities, which drive the subsequent evolution and govern the flame morphology observed at the onset of a detonation (Figure~\ref{fig:det_slice}). 

The models exhibit distinct flame morphologies that depend on both the progenitor central density and the ignition configuration. The single-kernel models (STD-o12r32, STD-o100r32, LOW-o12r32, HIGH-o12r32, HIGH-o80r32) broadly illustrate the combined effects of ignition offset and central density, which we discuss in turn.

Although STD-o12r32 and STD-o100r32 share the same progenitor central density ($\rho_\mathrm{c} \approx 2.2 \times 10^9\,\mathrm{g\,cm^{-3}}$), different central ignition kernel offsets in them place the kernels at distinct local densities at $t = 0$. The STD-o12r32 ignites with the WD center enclosed within the flame bubble, while the initial flame bubble of STD-o100r32 is at the lower density of $\sim 1.87 \times 10^9 \, \mathrm{g /cm^{3}}$. Since the critical bubble size required for the RT instability to overcome the inertia of the overlying fuel decreases with increasing local density, the more centrally ignited STD-o12r32 model develops a substantially thicker stem and suppresses RT onset relative to STD-o100r32. The larger-offset model (STD-o100r32), conversely, becomes RT unstable earlier, and the flame travels a greater distance before detonation is achieved. An analogous trend is observed in the high central density WD progenitors HIGH-o12r32 and HIGH-o80r32: the smaller-offset model develops a thicker, more columnar flame stem, while the larger-offset model undergoes earlier RT onset and greater pre-detonation flame propagation. 

Comparing models with similar ignition offsets across different progenitors, the HIGH-o12r32 model produces the largest flame bubble, spanning nearly 500 km before becoming RT unstable. This reflects the high-density progenitor's stronger gravitational field ($\sim 5.93 \times 10^9 \, \mathrm{{g} \, {cm^{-3}}}$), which increases the critical bubble size needed for the RT instability to overcome the inertia of the overlying fuel. At the opposite extreme, LOW-o12r32 model produces the thinnest stem and the least flame bubble growth prior to the RT onset, consistent with its lower-density progenitor providing weaker gravitational confinement. 

The STD-N100r32 model illustrates the effect of the multi-point ignition. The simultaneous ignition of 100 kernels effectively mimics a large, perturbed ignition surface, resulting in several RT plumes that rise rapidly towards the WD surface without forming a coherent single bubble. The individual flame bubbles remain small and the RT plumes are shorter compared to single-kernel models, as the distributed energy release suppresses the growth of any individual plume before the RT instability sets in across the entire burning surface. 


The deflagration phase converts C/O fuel into  iron-peak elements through nuclear burning, releasing energy into the surrounding medium. This energy deposition lowers the central density $\rho_\mathrm{c}$ and pre-expands the WD prior to tDDT, pushing the outer layers to progressively lower densities and increasing the volume of fuel below the IGE production threshold. The degree of pre-expansion is proportional to the total burned mass during the deflagration phase. 

The most significant pre-expansion is observed in the HIGH-o12r32 model, where $\rho_\mathrm{c}$ falls from $\sim 5.99 \times 10^9 \, \mathrm{{g} \, {cm^{-3}}}$ at ignition to $\sim 2.62 \times 10^9 \, \mathrm{{g} \, {cm^{-3}}}$ at the moment of tDDT -- a reduction of more than a factor of two. For the same progenitor, the larger-offset HIGH-o80r32 model burns less total mass during deflagration, resulting in substantially less pre-expansion, with $\rho_\mathrm{c}$ falling only to $\sim 4.13 \times 10^9 \, \mathrm{{g} \, {cm^{-3}}}$. The STD-density models undergo comparatively modest pre-expansion, reaching $\rho_\mathrm{c} \sim 2.00 \times 10^9 \, \mathrm{{g} \, {cm^{-3}}}$ at tDDT, while the LOW model experiences the least pre-expansion overall, with $\rho_\mathrm{c}$ declining to $ \rho_\mathrm{c}\sim 9.06 \times 10^9 \, \mathrm{{g} \, {cm^{-3}}}$ at detonation. The comparatively mild pre-expansion across most models results from the prompt nature of tDDT in our simulations: because the tDDT criterion is satisfied early in the deflagration phase, the flame has less time to burn fuel and release energy prior to detonation onset than in models employing a fixed density DDT trigger~\citep{seitenzahletal13b, leungnomoto18}.


In all models presented here, the detonation is prompt, occurring while the RT plumes are still rising through the WD interior. Figure~\ref{fig:det_slice} shows the state of each model at the moment of the first detonation trigger, with the green contours marking cells where $L/L_\mathrm{CJ} = 1$ and the blue contours indicating the leading edge of the flame front (flame progress variable $= 0.001$). Detonation is triggered when the green contours intersect with the blue, signaling that the local turbulent flame has exceeded the critical CJ volume required for pressure runaway. In the multi-point ignition model STD-N100r16, a total of three independent detonation points are triggered in the outermost RT plumes before the supersonic detonation wave sweeps through the WD interior. Figure~\ref{fig:det_slice} depicts the onset of the first such trigger. 

Once detonation ignition is achieved, the supersonic burning front rapidly consumes the remaining unburned fuel and leaves the WD gravitationally unbound. The burning products depend on the local fuel density at the time the detonation front arrives: fuel at densities above $\sim 1.7 \times 10^7 \, \mathrm{g/cm^3}$ is burned to iron-group elements (IGEs), while fuel below this threshold produces intermediate-mass elements (IMEs)~\citep{townsleyetal16}. The cumulative nuclear energy released across all models falls in the range $2.02$--$2.12 \times 10^{51} \,\mathrm{erg}$, increasing monotonically from the lowest to the highest central density progenitor. These values are in good agreement with previously published delayed-detonation models for near-$M_{Ch}$ progenitors~\citep{seitenzahletal13b,ohlmannetal14,leungnomoto18}, confirming that the tDDT mechanism produces energetically representative SNe~Ia explosions.


Once the supersonic detonation front disrupts the WD, the ejecta undergoes ballistic expansion as the pressure gradients become negligible relative to the bulk kinetic energy. The ejecta asymptotically approaches a state of homologous free expansion, in which the radial velocity of each fluid element is directly proportional to its distance from the center ($v_{r} \propto r$). This homologous state is a critical prerequisite for subsequent nucleosynthesis post-processing and radiative transfer calculations, as it confirms that the thermodynamic evolution of the ejecta is complete and the final ejecta velocity structure becomes frozen. Each simulation was continued until the homology criterion $\Delta \sim 0.1\%$ was satisfied before the tracer particle data were extracted for post-processing. 

\subsection{Ejecta Structure and Stratification}
\label{sec:structure}

\begin{figure}
    \centering
    \includegraphics[width=0.8\textwidth]{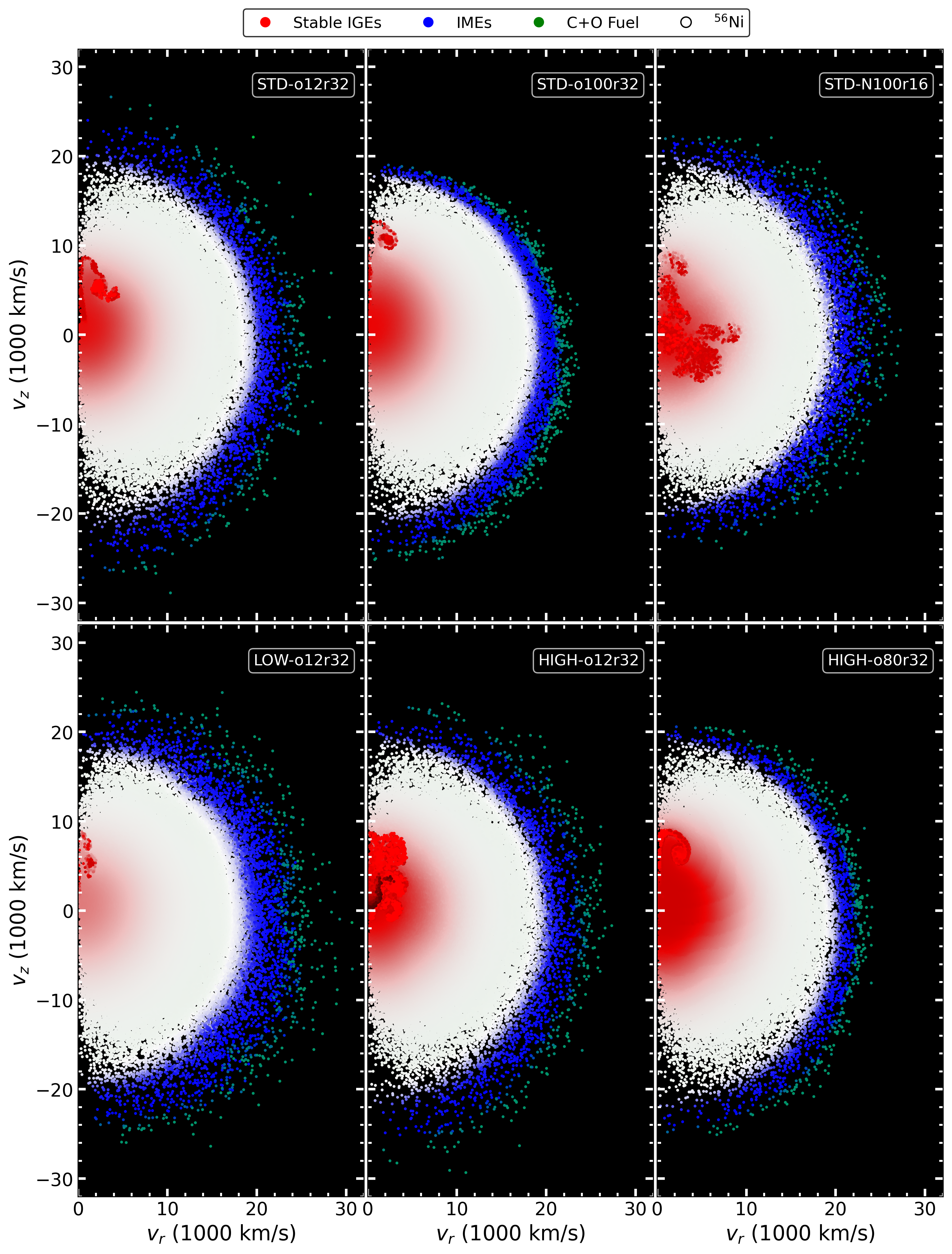}
    \caption{Tracer particles plotted in the z-r velocity space for all six models. The particles are colored on an RGB-W color scale, where red are stable IGEs ($22 \leq Z \leq 30$), green is unburned C$+$O, blue are IMEs (such as $^{28}$Si and $^{32}$S), and white are particles with $^{56}$Ni. For instance, a pink particle would represent stable IGEs and $^{56}Ni$ while a darker red or maroon particle would represent stable and unstable IGEs.}
    \label{fig:2d_structure}
\end{figure}

\begin{figure}
    \centering
    \includegraphics[width=0.8\textwidth]{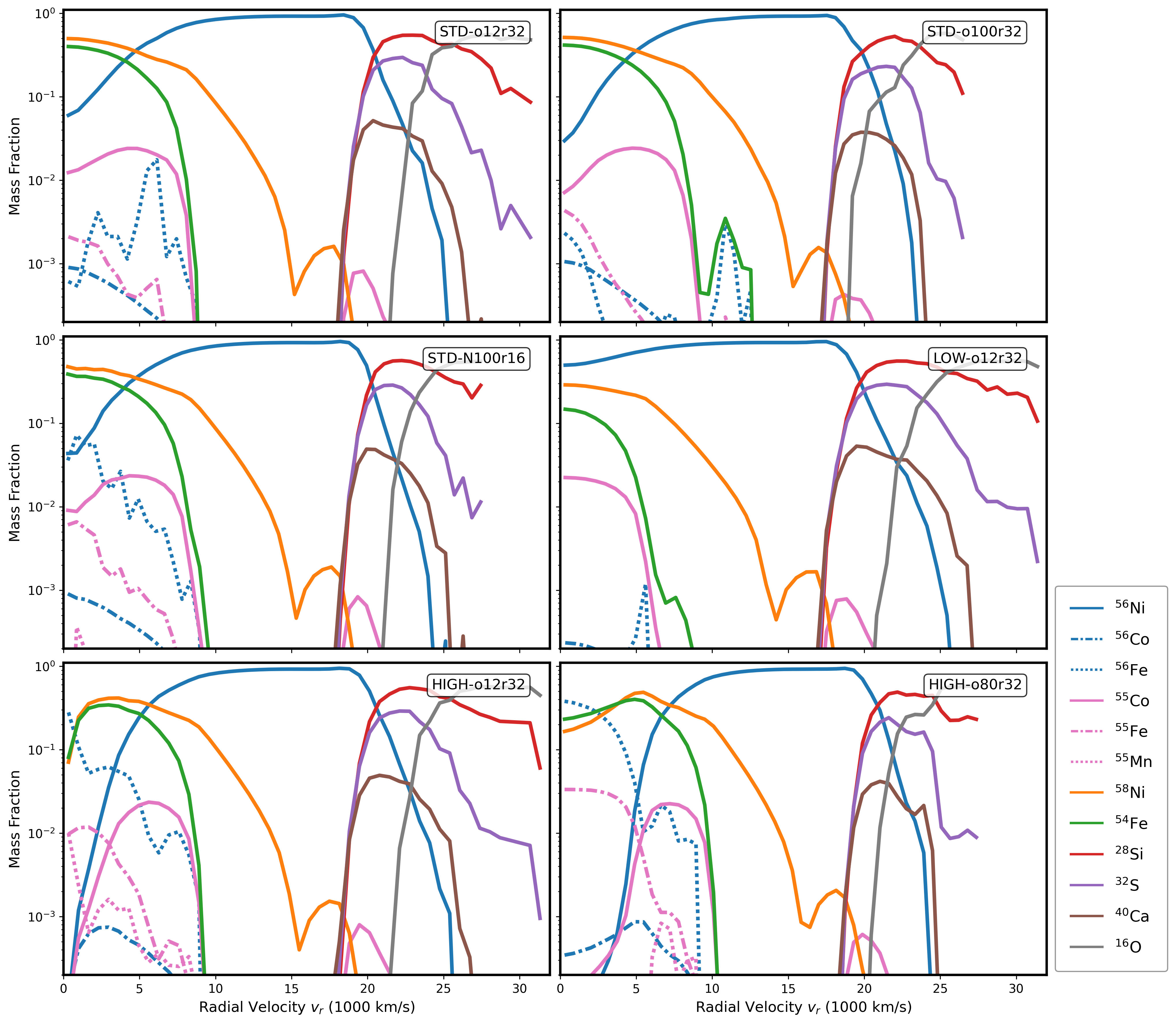}
    \caption{Velocity profiles for major isotopes from a radial average over the domain.}
    \label{fig:1d_structure}
\end{figure}

The final spatial distribution of the nucleosynthetic products is a direct kinetic record of the buoyant plume evolution and detonation physics described in Section~\ref{sec:ejecta}. As the ejecta enters homologous free expansion, chemical asymmetries generated during the deflagration phase are frozen in velocity space, establishing the primary link between the explosion's internal dynamics and its observed spectral signature~\citep{ropkeetal06, seitenzahletal13b}.

Figure~\ref{fig:2d_structure} shows the nucleosynthetic abundance stratification in the 2D cylindrical velocity space ($v_r, v_z$). Across the full model suite, a consistent global stratification emerges: a central concentration of stable IGEs (red) surrounded by a dominant $^{56}$Ni shell (white), an outer IME layer (blue), and unburned C$+$O fuel (green) confined to the highest velocities ($v \gtrsim 25{,}000\,\mathrm{km\,s^{-1}}$). This structure reflects the layered burning regimes of the detonation: complete NSE burning at high density in the interior, incomplete silicon and oxygen burning at the intermediate densities producing IMEs, and insufficient heating to consume carbon in the outermost layer. 

The stable IGE core (red) traces the subsonic deflagration phase, during which buoyant plumes of burning products facilitate electron capture that lowers the electron fraction ($Y_e$) prior to detonation, favoring the synthesis of neutron-rich stable NSE products~\citep{townsleyetal16}. Consequently, this core is concentrated in the wake of the ignition kernels rather than at the geometric center. Single-kernel offset models (STD-o12r32, STD-o100r32, LOW-o12r32, HIGH-o12r32, HIGH-o80r32) therefore show a pronounced asymmetry, with stable IGE production preferentially offset in the $+z$ direction corresponding to the direction of the plume rise. The distributed ignition model STD-N100r16, by contrast, exhibits more irregular and spatially ``smeared" stable IGE boundaries, as the detonation traverses the turbulent wakes of multiple simultaneously rising plumes. The volume of the stable IGE core scales monotonically with the progenitor central density, reflecting greater mass of fuel processed above the neutronization threshold ($\rho \gtrsim 10^9\,\mathrm{g\,cm^{-3}}$) in higher-density progenitors. 

Figure~\ref{fig:1d_structure} shows the radially averaged mass-fraction profiles for the isotopes most relevant to the radiative and spectroscopic evolution of SNe~Ia: the $^{56}$Ni decay chain ($^{56}$Ni, $^{56}$Co, $^{56}$Fe), the $A=55$ decay chain ($^{55}$Co, $^{55}$Fe, $^{55}$Mn), stable IGEs ($^{54}$Fe, $^{58}$Ni), dominant IMEs ($^{28}$Si, $^{32}$S, $^{40}$Ca), and unburned $^{16}$O. These species are selected as the primary drivers of radioactive heating, photospheric opacity, and spectral line formation~\citep{khokhlov91, seitenzahletal13b}. A key feature of the suite is that the abundance distributions are remarkably consistent across all models at high velocities ($v \gtrsim 15{,}000\,\mathrm{km\,s^{-1}}$), reflecting the insensitivity of the detonation-phase burning to the progenitor central density at lower values of this parameter. The models diverge significantly at low velocities ($\lesssim 7{,}000\,\mathrm{km\,s^{-1}}$), where differences in $\rho_c$ and the resulting electron fraction $Y_e$ at the time of tDDT directly govern the NSE freeze-out composition. The profiles are extracted at the epoch of homologous free expansion, at which point the spatial distributions of all species are frozen in the velocity space. 

The deepest ejecta ($v \lesssim 7{,}000\,\mathrm{km\,s^{-1}}$) corresponds to material processed at the highest densities and temperatures, where burning achieved complete NSE. Within this region, the innermost core ($v \lesssim 5{,}000\,\mathrm{km\,s^{-1}}$) is primarily composed of stable IGEs, namely $^{54}$Fe and $^{58}$Ni, whose relative abundances are highly sensitive to $\rho_c$ and $Y_e$ established during the deflagration phase. The zone between $5{,}000$ and $7{,}000\,\mathrm{km\,s^{-1}}$ marks the transition from this stable NSE core to the $^{56}$Ni-dominated shell, and this zone varies in width between the simulations depending on how much mass was burned at densities above the neutronization threshold.

In the high-density models (HIGH-o12r32 and HIGH-o80r32), the stable IGE core is markedly more pronounced. Efficient electron capture at $\rho_c \gtrsim 5 \times 10^9\,\mathrm{g\,cm^{-3}}$ suppresses $^{56}$Ni production at the lowest velocities and shifts the NSE freeze-out toward neutron-rich isotopes: $^{54}$Fe, $^{58}$Ni, and $^{56}$Fe are co-spatial and dominant within $v \lesssim 5{,}000\,\mathrm{km\,s^{-1}}$, constituting a ``normal" freeze-out regime in which the absence of abundant free $\alpha$-particles prevents $^{56}$Ni from dominating~\citep{woosley1986ApJ...301..601W}. The HIGH-o80r32 model, which experienced the least pre-expansion during the deflagration stage and, therefore, maintained higher burning densities at the time of detonation ignition, exhibits the most extended stable core of the suite. Beyond this stable core, $^{56}$Ni begins to dominate from $v \approx 5{,}000\,\mathrm{km\,s^{-1}}$ in the HIGH models. somewhat deeper than in the STD and LOW models where the $^{56}$Ni-dominated zone begins closer to $v \approx 3{,}000\,\mathrm{km\,s^{-1}}$.

At the opposite extreme, the LOW-o12r32 model shows a highly suppressed stable IGE core: its lower $\rho_c$ results in a $Y_e$ that remains close to 0.5 throughout the deflagration stage, favoring $^{56}$Ni production even at the lowest velocities. The STD-density models produce intermediate stable IGE signatures consistent with previous near-$M_\mathrm{Ch}$ delayed-detonation models of comparable central density~\citep{seitenzahletal13b}.

The distribution of the $A=55$ decay chain isotopes ($^{55}$Co, $^{55}$Fe, $^{55}$Mn) provides a particularly sensitive diagnostic of the neutronization history across the suite, as the relative abundances of these isobars are governed directly by $Y_e$ at the time of the NSE freeze-out~\citep{townsleyetal16, seitenzahletal13b}. 

In the standard central density models (STD-o12r32, STD-o100r32, and STD-N100r16), $^{55}$Co remains the dominant $A=55$ isotope throughout the inner ejecta, peaking near 5,000~km~s$^{-1}$. The $^{55}$Fe is detectable but subdominant, localized at the innermost core ($v \approx 2,000$~km~s$^{-1}$), indicating that neutronization in these models was insufficient to fully shift $A=55$ production from the proton-rich $^{55}$Co to the neutron-rich $^{55}$Fe. Notably, trace amounts of stable $^{55}$Mn are present only in the STD-N100r16 model at velocities $\approx$ 2,000~km~s$^{-1}$. The fragmented burning geometry of the 100-kernel ignition produces more spatially distributed high-density burning pockets, enabling direct synthesis of stable manganese that is otherwise suppressed in the single-kernel geometries. 

The low central density model (LOW-o12r32) shows the simplest $A=55$ chain distribution: $^{55}$Co peaks between $v \approx 0$ and $4{,}000\,\mathrm{km\,s^{-1}}$, with no detectable $^{55}$Fe or $^{55}$Mn. The low $\rho_c$ of this progenitor maintains electron fraction $Y_e \approx 0.5$ throughout the burning phase, preventing the neutronization required to synthesize the heavier $A=55$ isobars. This model is effectively ``neutron-poor" relative to the rest of the suite.

The high central density models (HIGH-o12r32 and HIGH-o80r32) exhibit a qualitatively distinct inversion. In both HIGH-o12r32 and HIGH-o80r32, $^{55}$Co is displaced to higher velocities between 6,000 and 7,000~km~s$^{-1}$, while the innermost ejecta are dominated by $^{55}$Fe and, in the HIGH-o80r32 model, by co-spatial $^{55}$Mn. This stratification is a direct consequence of the low-$Y_e$ environment created by efficient electron capture in the high-density core: neutron-rich species are synthesized in the deepest layers, pushing the proton-rich $^{55}$Co outward to the outer boundary of the NSE region. The HIGH-o80r32 model, with its limited pre-expansion, shows this inversion most prominently. The presence of $^{55}$Mn at the elevated expansion velocities, therefore, serves as a robust spectroscopic indicator of a high-density, single-degenerate progenitor, and offers a potential observational discriminant between the Chandrasekhar-mass and sub-Chandrasekhar-mass SNe~Ia scenarios~\citep{seitenzahletal13b,yamaguchietal15}. 

As the detonation wave propagates into lower-density regions of the expanding ejecta, the nuclear burning regime transitions from complete NSE to incomplete silicon and oxygen burning, producing a robust IME shell predominantly comprised of $^{28}$Si and $^{32}$S. This layer is remarkably consistent across the entire model suite, reflecting the insensitivity of the IME production to the deflagration-phase conditions once a detonation is established~\citep{townsleyetal16}. The IME shell is situated between the radioactive $^{56}$Ni core and the unburned $^{16}$O envelope, with the $^{28}$Si peak typically located between the velocities $\approx$ 20,000 and 25,000~km~s$^{-1}$. 

A distinct $^{40}$Ca peak near 20,000~km~s$^{-1}$ marks the transition from the $^{56}$Ni-dominated zone to the silicon-rich shell. This feature traces the oxygen-burning regime, where $\alpha$-captures on $^{28}$Si and $^{32}$S proceed until falling temperatures freeze the abundances~\citep{arnett69}. The only systematic variation with progenitor density is a modest outward shift in the IME inner boundary in the HIGH-density models: the $^{28}$Si peak shifts to velocities $\gtrsim$ 21,000~km~s$^{-1}$, reflecting the delayed transition from the IGE to IME burning due to a broader central high-density region. 

At the highest velocities ($v \gtrsim 25,000$~km~s$^{-1}$), the ejecta consist primarily of unburned $^{16}$O and $^{12}$C from the original 50/50 C/O WD. This unburned material is restricted to the outermost layers in all models, confirming that the tDDT mechanism triggers a detonation, which fully consumes the carbon-oxygen interior before expansion freezes nuclear reactions. The overall layered sequence, i.e., stable IGE core, $^{56}$Ni shell, IME layer, and unburned outer envelope, closely resembles the classical ``onion-shell" stratification of the 1D delayed-detonation models, such as W7~\citep{nomotoetal84} and its DDT successors~\citep{seitenzahletal13b}, providing confidence that the 3D tDDT framework employed here reproduces the essential nucleosynthetic architecture of the near-$M_\mathrm{Ch}$ SNe~Ia despite its fundamentally different detonation triggering mechanism. 

\subsection{Integrated Nucleosynthetic Yields}
\label{sec:yields}

\begin{figure}
    \centering
    \includegraphics[width=0.65\textwidth]{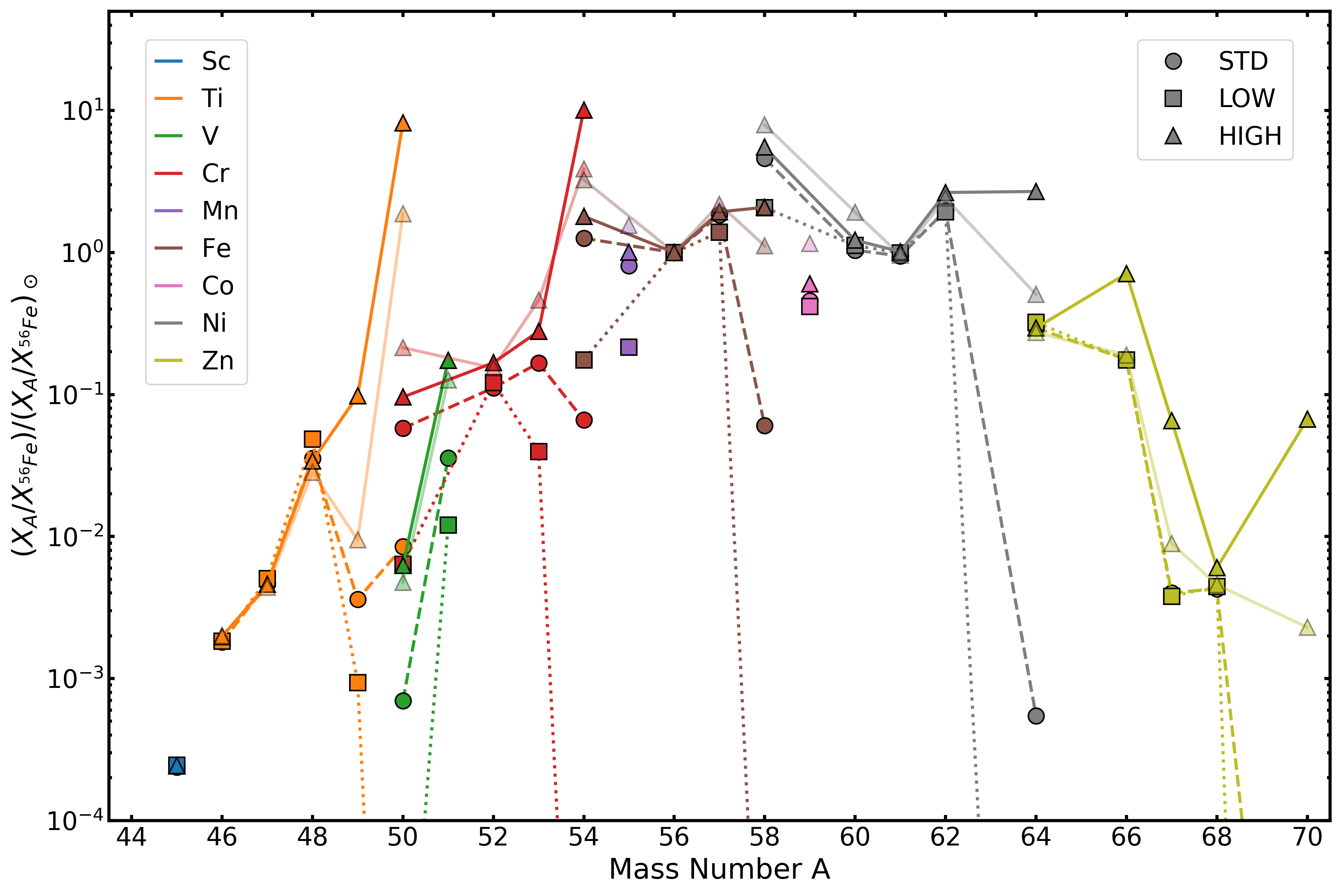}
    \caption{Isotopic abundance ratios of stable iron-peak isotopes to $^{56}$Fe normalized by the solar values from~\citet{lodders_2025}. Results are shown for four models spanning the progenitor density and ignition-offset parameter space:
    LOW-o12r32 (squares, dotted lines), STD-o12r32 (circles, dashed lines), HIGH-o12r32 (triangles, solid lines), and HIGH-o80r32
    (triangles, solid lines, lighter shading). Isotopes are color-coded by the element indicated in the legend.} 
    \label{fig:ratios}
\end{figure}

The total integrated nucleosynthetic yields for stable and radioactive isotopes across all six models are provided in Tables~\ref{table:stableYields} and \ref{table:radioactiveYields}, respectively. To isolate the effects of the progenitor central density and ignition geometry on the nucleosynthetic output, we focus our analysis on a representative subset: 1) the o12r32 configuration across all three progenitor density classes (LOW, STD, HIGH), which holds ignition geometry fixed while varying $\rho_c$, and 2) the HIGH-o80r32 model, which assesses the effect of the ignition offset at high density where the $\rho_c$ at tDDT varies most significantly with offset. Figure~\ref{fig:ratios} shows the resulting production factors of stable IGEs relative to the solar composition~\citep{lodders_2025}, which serves as the primary diagnostic for the neutronization history of the ejecta and highlights systematic deviations driven by these burning conditions. 

\startlongtable
\begin{deluxetable}{lcccccc}
\tablewidth{0pt}
\tablecaption{Stable nucleosynthetic yields ($M_\odot$) for all models. \label{table:stableYields}}
\tablehead{
\colhead{Isotope} & \colhead{STD-o12r32} & \colhead{STD-o100r32} & \colhead{STD-n100r32} & \colhead{LOW-o12r32} & \colhead{HIGH-o12r32} & \colhead{HIGH-o80r32} \\
\colhead{} & \multicolumn{6}{c}{($> 10^{-6}~M_{\odot}$)}
}
\startdata
$^{12}$C&1.35E-05&9.12E-06&4.56E-06&1.08E-05&8.92E-06&1.45E-05 \\
$^{16}$O&1.89E-03&3.71E-03&2.16E-03&3.29E-03&2.28E-03&2.32E-03 \\
$^{20}$Ne&1.08E-05&1.65E-05&5.30E-06&1.35E-05&1.14E-05&2.42E-05 \\
$^{24}$Mg&2.91E-04&8.03E-04&3.90E-04&5.42E-04&4.32E-04&5.31E-04 \\
$^{28}$Si&1.31E-02&1.60E-02&1.41E-02&2.17E-02&1.22E-02&9.34E-03 \\
$^{29}$Si&2.43E-06&3.30E-06&2.56E-06&3.95E-06&2.52E-06&2.11E-06 \\
$^{30}$Si&5.36E-06&4.39E-06&5.21E-06&8.20E-06&5.14E-06&3.77E-06 \\
$^{31}$P&2.55E-06&3.59E-06&2.72E-06&4.22E-06&2.63E-06&2.37E-06 \\
$^{32}$S&7.34E-03&8.17E-03&7.58E-03&1.22E-02&6.72E-03&4.76E-03 \\
$^{34}$S&2.55E-06&1.95E-06&2.56E-06&3.41E-06&2.43E-06&1.86E-06 \\
$^{35}$Cl&1.39E-06&1.38E-06&1.39E-06&1.97E-06&1.28E-06&1.02E-06 \\
$^{36}$Ar&1.62E-03&1.72E-03&1.63E-03&2.72E-03&1.49E-03&1.02E-03 \\
$^{39}$K&1.67E-06&1.67E-06&1.66E-06&2.10E-06&1.51E-06&1.29E-06 \\
$^{40}$Ca&1.62E-03&1.67E-03&1.60E-03&2.72E-03&1.48E-03&1.01E-03 \\
$^{44}$Ca&1.03E-05&9.74E-06&1.02E-05&1.38E-05&9.45E-06&7.40E-06 \\
$^{48}$Ti&6.40E-05&6.20E-05&6.27E-05&9.79E-05&5.84E-05&4.26E-05 \\
$^{51}$V&1.17E-05&1.10E-05&1.09E-05&4.41E-06&5.39E-05&3.45E-05 \\
$^{50}$Cr&3.63E-05&3.96E-05&3.72E-05&4.50E-06&5.72E-05&1.12E-04 \\
$^{52}$Cr&1.40E-03&1.30E-03&1.37E-03&1.71E-03&2.00E-03&1.61E-03 \\
$^{53}$Cr&2.43E-04&2.48E-04&2.52E-04&6.48E-05&3.85E-04&5.59E-04 \\
$^{55}$Mn&9.31E-03&9.80E-03&9.43E-03&2.80E-03&1.10E-02&1.49E-02 \\
$^{54}$Fe&7.65E-02&8.37E-02&7.68E-02&1.19E-02&1.04E-01&1.64E-01 \\
$^{56}$Fe&9.88E-01&9.63E-01&9.85E-01&1.11E+00&9.37E-01&8.22E-01 \\
$^{57}$Fe&4.28E-02&4.33E-02&4.28E-02&3.63E-02&4.26E-02&4.23E-02 \\
$^{59}$Co&1.43E-03&1.36E-03&1.48E-03&1.46E-03&1.77E-03&3.00E-03 \\
$^{58}$Ni&2.01E-01&2.14E-01&2.02E-01&1.02E-01&2.29E-01&2.89E-01 \\
$^{60}$Ni&1.81E-02&1.69E-02&1.89E-02&2.19E-02&2.02E-02&2.79E-02 \\
$^{61}$Ni&7.29E-04&7.12E-04&7.25E-04&8.60E-04&7.41E-04&6.13E-04 \\
$^{62}$Ni&4.94E-03&4.80E-03&4.84E-03&5.40E-03&6.27E-03&5.02E-03 \\
$^{63}$Cu&3.68E-05&2.99E-05&3.76E-05&4.33E-05&4.82E-05&3.36E-05 \\
$^{65}$Cu&5.34E-06&5.28E-06&5.30E-06&6.13E-06&9.74E-06&5.01E-06 \\
$^{64}$Zn&2.53E-04&2.27E-04&2.52E-04&3.12E-04&2.40E-04&1.96E-04 \\
$^{66}$Zn&8.96E-05&9.08E-05&8.96E-05&9.98E-05&3.43E-04&8.01E-05 \\
$^{68}$Zn&1.51E-06&1.16E-06&1.59E-06&1.76E-06&2.01E-06&1.34E-06 \\
$^{70}$Ge&2.39E-06&2.43E-06&2.40E-06&2.59E-06&2.29E-06&2.07E-06 \\
\enddata
\end{deluxetable}


\startlongtable
\begin{deluxetable}{lcccccc}
\tablewidth{0pt}
\tablecaption{Radioactive nucleosynthetic yields ($M_{\odot}$) for all models. \label{table:radioactiveYields}}
\tablehead{
\colhead{Isotope} & \colhead{STD-o12r32} & \colhead{STD-o100r32} & \colhead{STD-n100r32} & \colhead{LOW-o12r32} & \colhead{HIGH-o12r32} & \colhead{HIGH-o80r32} \\
\colhead{} & \multicolumn{6}{c}{($> 10^{-6}~M_{\odot}$)}
}
\startdata
$^{30}$P&3.48E-06&3.19E-06&3.79E-06&5.78E-06&3.31E-06&2.89E-06 \\
$^{31}$S&1.50E-06&1.95E-06&1.42E-06&2.30E-06&1.58E-06&1.14E-06 \\
$^{44}$Ti&1.03E-05&9.74E-06&1.02E-05&1.38E-05&9.45E-06&7.40E-06 \\
$^{48}$Cr&6.39E-05&6.20E-05&6.26E-05&9.79E-05&5.81E-05&4.24E-05 \\
$^{51}$Mn&1.06E-05&1.10E-05&1.03E-05&4.37E-06&1.15E-05&1.15E-05 \\
$^{52}$Fe&1.29E-03&1.30E-03&1.26E-03&1.71E-03&1.20E-03&9.42E-04 \\
$^{53}$Fe&2.27E-04&2.42E-04&2.23E-04&6.38E-05&2.45E-04&2.50E-04 \\
$^{55}$Fe&2.30E-04&1.89E-04&4.45E-04&2.06E-05&1.07E-03&4.86E-03 \\
$^{55}$Co&9.06E-03&9.61E-03&8.98E-03&2.78E-03&9.63E-03&9.83E-03 \\
$^{56}$Co&1.44E-04&1.60E-04&1.45E-04&3.51E-05&2.00E-04&3.18E-04 \\
$^{57}$Co&1.43E-04&1.12E-04&2.85E-04&1.53E-05&6.98E-04&3.31E-03 \\
$^{56}$Ni&9.85E-01&9.63E-01&9.80E-01&1.11E+00&9.25E-01&7.88E-01 \\
$^{57}$Ni&4.26E-02&4.31E-02&4.25E-02&3.63E-02&4.17E-02&3.89E-02 \\
$^{59}$Ni&2.17E-04&2.25E-04&2.88E-04&5.92E-05&5.45E-04&1.97E-03 \\
$^{58}$Cu&2.46E-04&1.91E-04&2.20E-04&2.93E-04&2.33E-04&1.63E-04 \\
$^{59}$Cu&1.20E-03&1.14E-03&1.19E-03&1.40E-03&1.12E-03&9.78E-04 \\
$^{60}$Cu&6.71E-05&7.53E-05&8.32E-05&9.58E-05&6.04E-05&7.23E-05 \\
$^{61}$Cu&9.59E-06&1.12E-05&1.19E-05&1.18E-05&9.44E-06&1.36E-05 \\
$^{60}$Zn&1.73E-02&1.65E-02&1.72E-02&2.18E-02&1.60E-02&1.33E-02 \\
$^{61}$Zn&7.15E-04&7.01E-04&7.09E-04&8.48E-04&6.70E-04&5.74E-04 \\
$^{62}$Zn&4.77E-03&4.80E-03&4.76E-03&5.40E-03&4.49E-03&4.00E-03 \\
$^{63}$Zn&3.67E-06&3.64E-06&4.09E-06&4.31E-06&3.46E-06&3.67E-06 \\
$^{63}$Ga&3.30E-05&2.63E-05&3.34E-05&3.90E-05&3.18E-05&2.64E-05 \\
$^{64}$Ga&3.60E-06&3.79E-06&4.41E-06&4.97E-06&3.31E-06&3.81E-06 \\
$^{64}$Ge&2.49E-04&2.24E-04&2.48E-04&3.07E-04&2.36E-04&1.91E-04 \\
$^{65}$Ge&4.90E-06&4.79E-06&4.76E-06&5.58E-06&4.63E-06&3.90E-06 \\
$^{66}$Ge&8.95E-05&9.08E-05&8.96E-05&9.98E-05&8.50E-05&7.62E-05 \\
$^{68}$Se&1.43E-06&1.08E-06&1.49E-06&1.66E-06&1.43E-06&1.12E-06 \\
$^{70}$Se&2.39E-06&2.43E-06&2.40E-06&2.59E-06&2.29E-06&2.07E-06 \\
\enddata
\end{deluxetable}


The total mass of radioactive $^{56}$Ni varies significantly across the suite and represents the primary driver of peak luminosity through the $^{56}$Ni $\to$ $^{56}$Co $\to$ $^{56}$Fe decay chain~\citep{muellerarnett82}. We find a clear monotonic decline of the $^{56}$Ni mass with increasing progenitor central density: LOW-o12r32 synthesizes $\sim 1.11$\,M$_\odot$, the STD models produce $\sim 0.98\,M_\odot$, and the HIGH models yield $\sim 0.93\,M_\odot$ (HIGH-o12r32) and $\sim 0.79\,M_\odot$ (HIGH-o80r32). This decline is a direct consequence of enhanced electron capture rates at high densities ($\rho \gtrsim 10^9 \text{ g cm}^{-3}$), which convert $^{56}$Ni precursors into stable, neutron-rich isotopes during the deflagration phase~\citep{nomoto1997Sci...276.1378N}. This trend is illustrated in Fig.~\ref{fig:ratios}, which shows that stable nickel isotopes (grey markers) increase systematically with progenitor density. 

The $^{56}$Ni yields across all models ($0.79$--$1.11\,M_\odot$) substantially exceed the $\sim 0.4$--$0.6\,M_\odot$  characteristic of normal SNe Ia \citep{stritzingeretal06}. This is a consequence of the prompt tDDT: because detonation is triggered early while the WD remains compact and only modestly pre-expanded, the burning front propagates through high-density fuel and drives nearly complete incineration to NSE, producing large amounts of $^{56}$Ni and suppressing the IME yields that would otherwise reduce the nickel budget.

High-density burning leaves a distinct imprint on the stable IGE composition (Figure~\ref{fig:ratios}). The yield of $^{54}$Fe -- a sensitive tracer of neutronization -- increases by nearly an order of magnitude from the LOW-o12r32 ($\sim 0.012$\,M$_\odot$) to the HIGH models ($0.10$--$0.16$\,M$_\odot$), as the central density of the LOW model falls just below the threshold for efficient electron capture. Similarly, the monoisotopic odd-Z element $^{55}$Mn rises from $2.8 \times 10^{-3}$\,M$_\odot$  in LOW-o12r32 to $\sim 1.5 \times 10^{-2}$\,M$_\odot$ in HIGH-o80r32, reinforcing its utility as a secondary diagnostic of the progenitor central density~\citep{seitenzahletal13b}. Stable $^{58}$Ni also increases monotonically, from $0.10\,M_\odot$ (LOW) to $0.29\,M_\odot$ (HIGH-o80r32), reflecting the broader shift of the NSE freeze-out toward neutron-rich species at high $\rho_c$.

A comparison of the two high-density models reveals that nucleosynthetic output depends not only on the density at detonation, but also on the specific thermodynamic history of the deflagration phase. The HIGH-o80r32 model retains the highest central density at the time of detonation ($\rho_{\text{tDDT}}$) owing to its limited pre-expansion, producing the largest yields of bulk neutron-rich species in the suite: $^{54}$Fe (0.16\,M$_{\odot}$) and $^{58}$Ni (0.29\,M$_{\odot}$). The HIGH-o12r32 model, despite detonating at a somewhat lower density, produces higher yields of the extreme neutron-rich tracers $^{54}$Cr ($A=54$), $^{50}$Ti ($A=50$), and $^{48}$Ca ($A=48$). These rate isotopes are preferentially synthesized in the more centrally located deflagration product, which is more extensive in the centrally ignited o12 geometry and thus which undergoes prolonged electron captures during the subsonic burning. In contrast, the o80 offset geometry leaves the WD center as unburned fuel until the prompt detonation front arrives, favoring the bulk neutron-rich IGEs over the extreme neutronization tracers.

The overproduction of $^{54}$Cr and $^{50}$Ti in the high-density models, with yields reaching $\sim 10^{-3}\,M_\odot$, places a strong constraint on the event rate. If high-density Chandrasekhar-mass explosions ($\rho_c \geq 5.5 \times 10^9\,\mathrm{g\,cm^{-3}}$) were a common SNe~Ia channel, the resulting galactic chemical evolution would be distorted beyond observational bounds for these elements~\citep{iwamoto1999nucleosynthesis}. 
These results are consistent with the interpretation that high-density near-$M_\mathrm{Ch}$ events represent a minor contributor to the overall SNe~Ia rate.  
The high-density models also activate deep neutronization channels unavailable at lower densities, producing $^{58}$Fe ($6.2 \times 10^{-3}$\,M$_\odot$) and $^{64}$Ni at levels negligible in the low-density model, in agreement with earlier high-density delayed-detonation studies~\citep{seitenzahletal13b, leungnomoto18}.

Several radioactive isotopes provide explosion diagnostics at late times. Radioactive $^{55}$Fe scales strongly with the progenitor central density, ranging from $\sim 2 \times 10^{-5}\,M_\odot$ (LOW-o12r32)
to $\sim 4.9 \times 10^{-3}\,M_\odot$ (HIGH-o80r32). Its electron-capture decay to $^{55}$Mn produces a 5.9 keV X-ray line whose flux scales with the $^{55}$Fe yield, offering a potential spectroscopic diagnostic in the nearby young supernova remnants~\citep{seitenzahletal15}. 

In contrast, $^{44}$Ti yields are consistently low across the suite ($\sim 7$--$14 \times 10^{-6}\,M_\odot$), substantially below the yields expected from the helium-shell detonation scenarios ($\sim 10^{-4}$--$10^{-3}\,M_\odot$)~\citep{The2006, gronowetal2021}. This difference confirms that none of our models are consistent with the sub-Chandrasekhar double-detonation events, and that the $^{44}$Ti yields alone can in principle discriminate between these explosion pathways in resolved remnants. 

The ratio of stable to radioactive nickel ($M_{^{58}\text{Ni}} / M_{^{56}\text{Ni}}$) increases monotonically with the progenitor density across the suite, from $\sim\!0.092$ (LOW-o12r32) to $\sim\!0.37$ (HIGH-o80r32). This ratio is accessible through late-time nebular spectroscopy, where the stable $^{58}$Ni contributes to forbidden [Ni~\textsc{ii}] emission while $^{56}$Ni governs the bolometric light curve, making it a robust discriminant of the progenitor central density in the well-observed late-time spectra~\citep{blondinetal17}.

The extent of nuclear burning is characterized by the remarkably low yields of the IMEs and unburned fuel across all models. The integrated yields of $^{28}$Si (0.009--0.022\,M$_\odot$), $^{32}$S (0.005--0.012\,M$_\odot$), $^{36}$Ar ($0.001$--$0.003\,M_\odot$), $^{40}$Ca ($0.001$--$0.003\,M_\odot$), and unburned $^{16}$O ($\sim 0.002$--$0.004\,M_\odot$) are an order of magnitude below the $\sim 0.2$--$0.4\,M_\odot$ of silicon characteristic of standard 1D delayed-detonation models~\citep{nomotoetal84,seitenzahletal13b}. These reduced IME yields are a direct consequence of the prompt nature of the tDDT trigger: because detonation is initiated early, while the WD remains compact and the fuel density is still high, the supersonic burning front encounters material above the complete-burning threshold ($\rho \gtrsim 1.7 \times 10^7\,\mathrm{g\,cm^{-3}}$~\citep{townsleyetal16}) throughout most of the stellar volume, driving nearly complete incineration to NSE rather than the incomplete silicon-burning regime that produces IMEs. A modest positive correlation exists between the IME yields and $\rho_c$: the low-density model produces the most $^{28}$Si ($0.022\,M_\odot$) because its lower $\rho_c$ leads to greater pre-expansion before tDDT, reducing the fraction of fuel above the NSE threshold at detonation. 

\subsection{Spectral Analysis}
\label{sec:spectra}

\begin{deluxetable*}{lccccc}
\tablewidth{0pt}
\tablecaption{SNID cross-correlation results for synthetic spectra near peak brightness using the BSNIP template set, with template ages constrained to $\pm 3$~days relative to maximum brightness. For each model we list the two highest-ranked matches returned by SNID, along with the correlation statistics (\texttt{rlap}), the best-fit redshift, and the best-fit template age. The reported redshift is the global wavelength shift favored by SNID and is not interpreted as a cosmological redshift for the synthetic spectra. \label{table:snid_rlap}}
\tablehead{
\colhead{Model} & \colhead{Rank} & \colhead{Template} & \colhead{Type} &
\colhead{\texttt{rlap}} & \colhead{Age (days)}
}
\startdata
\multirow{2}{*}{STD-o12r32} & 1 & SN~99aa & Ia-99aa & 6.90 & $+2.70$ \\
& 2 & SN~99aa & Ia-99aa & 5.71 & $-0.60$ \\
\hline
\multirow{2}{*}{STD-o100r32} & 1 & SN~99aa & Ia-99aa & 8.04 & $+2.70$ \\
& 2 & SN~99aa & Ia-99aa & 6.63 & $-0.60$ \\
\hline
\multirow{2}{*}{STD-n100r16} & 1 & SN~99aa & Ia-99aa & 6.88 & $+2.70$ \\
& 2 & SN~99aa & Ia-99aa & 6.19 & $-0.60$ \\
\hline
\multirow{2}{*}{LOW-o12r32} & 1 & SN~99aa & Ia-99aa & 5.95 & $+2.70$ \\
& 2 & SN~99aa & Ia-99aa & 5.07 & $-0.60$ \\
\hline
\multirow{2}{*}{HIGH-o12r32} & 1 & SN~99aa & Ia-99aa & 6.12 & $+2.70$ \\
& 2 & SN~99aa & Ia-99aa & 5.44 & $-0.60$ \\
\hline
\multirow{2}{*}{HIGH-o80r32} & 1 & SN~99aa & Ia-99aa & 8.62 & $+2.70$ \\
& 2 & SN~99aa & Ia-99aa & 6.92 & $-0.60$ \\
\enddata
\end{deluxetable*}


We assess the observational fidelity of all six models by comparing their synthetic spectra near peak brightness to a library of observed SNe Ia events using the \texttt{SNID} code~\citep{blondintonry07}. \texttt{SNID} performs a cross-correlation in the logarithmic wavelength space between the input spectrum and each library template after removing the pseudo-continuum, thereby emphasizing line features over broadband color. 

For each model, we classify the synthetic spectrum at peak brightness against the Berkeley Supernova Ia Program (BSNIP, \citet{silvermanetal12_bsnip}) template dataset distributed with \texttt{SNID}, restricting the template epochs to within $\pm 3$~days of peak brightness to reduce phase-driven spectral diversity. The two highest rank matches for each model are summarized in Table~\ref{table:snid_rlap}. 

All six models produce the highest-ranked matches to SN~1999aa template (subtype Ia-99aa), with best-match $\texttt{rlap}$ values spanning $5.95-8.62$ across the suite and $6.90-8.04$ among the three standard-density models. The inferred best-match phases cluster near peak brightness (within the imposed $\pm 3$~day window), confirming that the agreement is driven by the genuine morphological similarity in the line-pattern structure rather than a phase mismatch. The systematic identification of all six models as SN~1999aa-like, regardless of the ignition geometry or progenitor central density, indicates that the tDDT mechanism robustly produces this spectroscopic subtype across the parameter space explored here.

The variation of \texttt{rlap} across the suite does not follow a simple monotonic trend with the progenitor density. The highest match quality is achieved by STD-o100r32 ($\texttt{rlap} = 8.04$) and HIGH-o80r32 ($\texttt{rlap} = 8.62$), while HIGH-o12r32 ($\texttt{rlap} = 6.12$) yields a lower score than both STD single-kernel models despite its higher progenitor density. This non-monotonic behavior suggests that the spectral match quality is governed not solely by $\rho_c$, but rather by the combined effects of the ignition offset, pre-expansion history, and the resulting velocity structure of the iron-group-rich ejecta. In particular, the HIGH-o80r32 model's high \texttt{rlap} is consistent with its higher $\rho_\mathrm{tDDT}$ and limited pre-expansion, which produces a more concentrated, hotter photosphere at peak that more closely reproduces the Fe~\textsc{iii}-dominated features of SN~1999aa.

SN~1999aa-like events occupy a distinct region within the spectroscopic classification of SNe Ia, characterized near maximum light by the prominent iron-group absorption (Fe~\textsc{iii}-dominated) at blue wavelengths and comparatively weak intermediate-mass-element absorption features, especially Si~\textsc{ii} $\lambda 6355$~\citep{filippenkoetal92, ruizlapuenteetal92}. While SN~1999aa-like events are often grouped with the broader 91T family, they are spectroscopically distinct: SN~1991T-like events show almost complete suppression of Si~\textsc{ii} and Ca~\textsc{ii} near maximum, whereas SN~199aa-like events retain detectable but shallow Si~\textsc{ii} absorption, representing an intermediate class between 91T-like and branch-normal SNe~Ia~\citep{benettietal05,foleyetal09}.

Within thermonuclear-ejecta models, such ``shallow silicon'' behavior characteristic of this subclass can arise from two physical mechanisms. The first one is ionization: higher photospheric temperatures shift silicon and calcium to higher ionization stages (e.g., Si~\textsc{iii}, Ca~\textsc{iii}), suppressing the optical Si~\textsc{ii} and Ca~\textsc{ii} absorption lines even when the underlying IME mass is non-negligible~\citep{filippenkoetal97}. The second one is abundance: low IME yields caused by nearly complete incineration to NSE reduce the optical depth in Si~\textsc{ii} directly. Both mechanisms operate simultaneously in our tDDT models: prompt detonation, which is triggered while the WD remains compact, produces high photospheric temperature and suppresses IME production. The enhanced presence of Fe~\textsc{iii} features near maximum light is further facilitated by the iron-group material transported to the outer ejecta through turbulent mixing during the deflagration phase~\citep{simetal13,seitenzahletal13b}.

\begin{figure*}[t]
\centering
\includegraphics[width=0.8\textwidth]{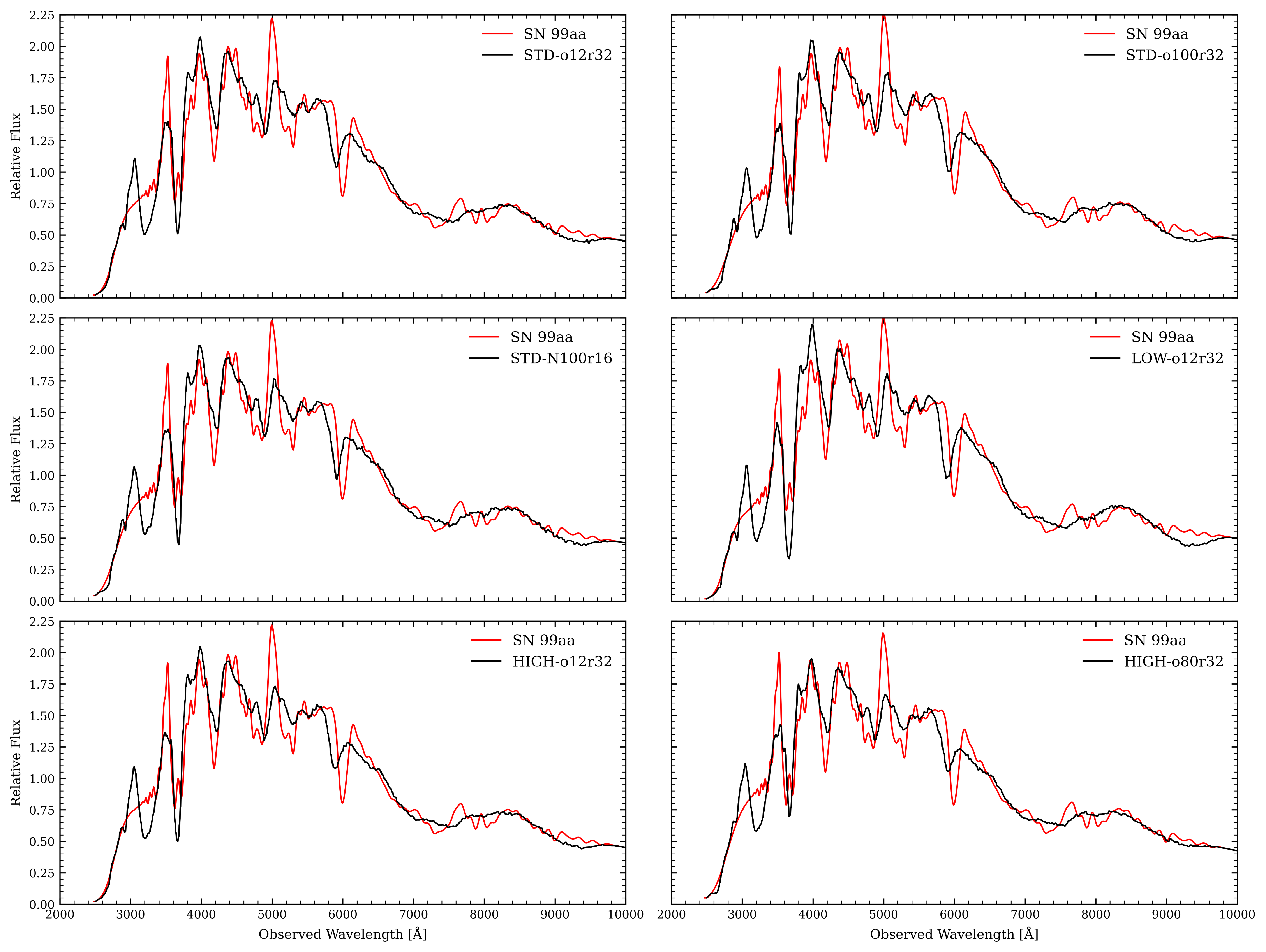}
\caption{Comparison of synthetic spectra near maximum light (black) with the highest-ranked SNID template match (red) for the six models analyzed here.
}
\label{fig:spectra_comparison}
\end{figure*}


Figure~\ref{fig:spectra_comparison} shows that the synthetic spectra reproduce the overall line-pattern morphology of the top-ranked SN~1999aa template from the near UV to the near IR ($\sim 2500$--$10{,}000$~\AA), including the dense iron-group absorption complex at $\lambda \lesssim 5000$~\AA\ and the shallow Si~\textsc{ii}~$\lambda6355$ feature typical of this subclass near peak~\citep{filippenkoetal97}. Across the model suite, the match quality is broadly uniform, implying that the predicted maximum-light spectral appearance is relatively insensitive to the specific ignition configuration and central-density variations at the level probed by the \texttt{SNID} cross-correlation. This is consistent with the general finding that multiple progenitor and explosion parameters can map onto similar photospheric spectra once the temperature--ionization structure is fixed~\citep{kruegeretal12,simetal13}. 

A closer inspection of the individual spectral features reveals several systematic residuals. The iron-group absorption complex near 4000-5000~\AA\ is stronger in the SN~1999aa template than in the synthetic spectra across all models, suggesting a modest underestimate of the iron-group opacity or mixing in the outer ejecta at this epoch. The Si~\textsc{ii} $\lambda 6355$ feature shows a systematic blue-shift in the synthetic spectra relative to the SN~99aa template, indicative of a higher photospheric velocity in the models than in the actual SN~1999aa explosion. 

Using the pseudo-equivalent-width diagnostics of \citet{benettietal05}, the weakness of the $\lambda 6355$ feature combined with a stronger absorption near 5900~\AA\ places the synthetic spectra within the shallow-silicon branch of the Branch classification diagram~\citep{benettietal05}, consistent with the \texttt{SNID} identification and with the low IME yields. Template content below $\sim 3500$~\AA\ provides less discriminating power due to the increased Monte Carlo noise in the UV and reduced template coverage at these wavelengths in the BSNIP dataset~\citep{silvermanetal12_bsnip}.



Our model set spans a factor of six in central ignition density ($1.0$--$6.0 \times 10^9$~g~cm$^{-3}$), single-point ignition offsets from 12 to 100~km, and qualitatively distinct ignition topologies (single-point versus 100-point multi-point ignition). Despite these substantial variations in initial conditions, every model returns SN~1999aa as both its first- and second-ranked \texttt{SNID} template match, with spectra near peak that are virtually indistinguishable across the model set.

\section{Conclusion and discussion}
\label{chap:conclusion}
The transition from a subsonic deflagration to a supersonic detonation in a 
near-Chandrasekhar-mass WD has been treated as a parametrization 
for three decades. Since \citet{khokhlovetal97} and \citet{niemeyerwoosley97}, 
the standard practice has been to trigger a detonation at an empirically tuned 
critical density. Such prescription works. It is also not a mechanism. The 
physical process responsible for DDT in WDs has remained unidentified.

The missing physics of detonation initiation has been provided by the turbulence-driven DDT mechanism of \citet{poludnenkoetal19}, validated in laboratory detonation experiments. A 
deflagration need not reach the detonation speed to initiate a pressure 
runaway. It only needs to exceed the Chapman-Jouguet (CJ) deflagration speed -- a 
threshold far lower, and physically well-defined. We have applied this 
framework, for the first time, to the interior of an exploding WD. 
Six 3D full-star \texttt{FLASH} simulations were coupled to \texttt{Torch} 
nucleosynthesis post-processing and to \texttt{SuperNu} radiative transfer. 
The suite spans central densities from $\rho_c \approx 1.0 \times 10^9$ to 
$6.0 \times 10^9\,\mathrm{g\,cm^{-3}}$ and qualitatively distinct ignition 
geometries, including a 100-kernel distributed ignition, spanning the widest
range of possible WD progenitors and ignition geometries.

In these simulations, detonation is not triggered at 
a tuned transition density. Instead, it is initiated when local turbulence first satisfies the CJ 
threshold ($U_l \gtrsim U_\mathrm{CJ}$ on the scale $L_\mathrm{CJ}$) -- a 
condition, which depends only on the local turbulent conditions and fuel properties and which is, therefore, reached 
at a nearly common pre-expansion state across our model set, regardless of 
the path taken to get there. The detonation that follows propagates through 
a similarly structured WD in each case, producing the outer 
intermediate-mass-element layer, which dominates the photospheric spectrum near 
peak brightness. In this picture, the spectroscopic uniformity of near-$\Mch$ WD progenitors
is not a coincidence requiring fine-tuning, but a mechanistic consequence of 
a turbulence-controlled trigger that gives rise to similar detonation configurations from the
diverse WD progenitors and ignition initial conditions. This provides a natural foundation for the kind of  ignition-insensitive detonation outcomes implicit in the empirical 
standardizability of SNe~Ia.

The tDDT criterion is evaluated locally and instantaneously from the resolved flow. It depends 
only on the local turbulent conditions and fuel properties, not on the global state of the WD. Different ignition histories therefore 
drive the system along different trajectories, but all reach the CJ 
threshold at a nearly common pre-expansion state. The detonation that 
follows propagates through a similarly structured WD in each case. 
The spectroscopic outcome reflects this convergence: synthetic spectra near 
peak brightness are virtually indistinguishable across the suite, and every model 
returns the same \texttt{SNID} first- and second-ranked template match. The tDDT 
mechanism delivers a turbulence-controlled trigger that selects a common 
detonation configuration from a diverse set of initial conditions.

This convergence has specific consequences for the structure of tDDT-driven SN Ia 
ejecta. Detonation is prompt. The deflagration phase is less vigorous than 
in the parameterized DDT models, and central densities at the moment of 
detonation are correspondingly higher. The stable IGE core is more 
pronounced in the high-density models, with $^{55}$Fe and $^{55}$Mn 
displacing $^{55}$Co to higher velocities and dominating the innermost 
ejecta -- a qualitative inversion of the $A=55$ chain that is absent in 
the low-density model. Integrated $^{56}$Ni masses span $0.79$--$1.11\,
M_\odot$. IME yields are suppressed by an order of magnitude relative to the
standard delayed-detonation models. The $^{44}$Ti yields remain at 
$\sim 10^{-5}\,M_\odot$ across the entire suite, well below the levels 
expected from the sub-Chandrasekhar double-detonation channels.

It is tempting, but likely premature to ascribe 91T/99aa-like events as robust 
signatures of near-$\Mch$ WD progenitors, given that these models still lack
the magnetic field (see below), which may play a key role in determining the ultimate
outcome of the deflagration phase in this channel. However, it is clear that turbulence-driven 
detonation in near-$M_\mathrm{Ch}$ white dwarfs, applied without tuning to 
match any spectral observable, produces a narrow and reproducible 
spectroscopic outcome at peak in purely hydrodynamical simulations. This is the kind of ignition-insensitive 
behavior implicit in the empirical standardizability of SNe~Ia. Whether 
tDDT $\Mch$ WD progenitors contribute substantially to the observed SN~Ia population, and to 
which branch of the spectroscopic sequence, are questions for further work.

The largest open question concerns magnetic fields. Our simulations are 
purely hydrodynamic. They presume that Rayleigh-Taylor-driven turbulence 
develops to the point of satisfying the CJ criterion without modification 
by magnetic stresses. The pre-explosion magnetic field of a near-$M_
\mathrm{Ch}$ white dwarf is poorly constrained. Magnetic tension on the 
relevant scales could suppress the turbulent cascade, raise the effective 
threshold for $U_l \geq U_\mathrm{CJ}$, or prevent the threshold from 
being reached at all. The consequences span a wide range: more vigorous 
pre-expansion shifting the explosion toward sub-luminous outcomes, or 
complete failure to detonate, leaving a bound or partially bound remnant 
rather than a normal SN~Ia. Whether tDDT operates as cleanly in MHD as 
it does in pure hydrodynamics is a central physics question for the 
next generation of tDDT SNe Ia simulations.

\begin{acknowledgments}

R.T.F. acknowledges support by the NSF award AST-2511516, and K.P., A.D., R.T.F., and A.Y.P. ackowledge support by the NASA Astrophysics Theory Program award 80NSSC22K0630. This work used Stampede 3 supercomputer at University of Texas at Austin's Texas Advanced Computing Center through allocation TG-AST100038 from the Advanced Cyberinfrastructure Coordination Ecosystem: Services \& Support (ACCESS) program, which is supported by the NSF grants \#2138259, \#2138286, \#2138307, \#2137603, and \#2138296 \citep{boerner2023access}. 

\end{acknowledgments}
\begin{contribution}

K.P. and A.D. performed the high-resolution \texttt{FLASH} simulations presented here, analyzed the results, and led the writing of the manuscript. M.U. implemented the tDDT mechanism in FLASH, conducted the initial test simulations that first demonstrated prompt detonations, and contributed to the analysis pipeline. M.U. and C.B. both performed early pathfinder simulations, and contributed to the analysis pipeline. A.P. and V.G. both led the incorporation of the tDDT mechanism into the 3D hydro simulations. R.T.F. supervised the work of K.P., A.D., M.U., and C.B. for the complete duration of the project and provided important feedback to K.P. and A.D. while writing the manuscript. 

\end{contribution}

\section{Software and third party data repository citations} \label{sec:cite}

\software{Flash 4.0 \citep{fryxell2000flash},  Python programming language \citep{vanrossumdeboer91}, Yt \citep{turk_2011}, Numpy \citep{vanderwaltetal11}, Matplotlib \citep{hunter07}, Jupyter/Ipython \citep{perezgranger07}, Scipy \citep{2020SciPy-NMeth}}






\begin{thebibliography}{}
\expandafter\ifx\csname natexlab\endcsname\relax\def\natexlab#1{#1}\fi
\providecommand{\url}[1]{\href{#1}{#1}}
\providecommand{\dodoi}[1]{doi:~\href{http://doi.org/#1}{\nolinkurl{#1}}}
\providecommand{\doeprint}[1]{\href{http://ascl.net/#1}{\nolinkurl{http://ascl.net/#1}}}
\providecommand{\doarXiv}[1]{\href{https://arxiv.org/abs/#1}{\nolinkurl{https://arxiv.org/abs/#1}}}

\bibitem[{W.~D. {Arnett}(1969){Arnett}}]{arnett69}
{Arnett}, W.~D. 1969, \bibinfo{title}{{A Possible Model of Supernovae:
  Detonation of $^{12}$C},} \apss, 5, 180, \dodoi{10.1007/BF00650291}

\bibitem[{C. {Ashall} {et~al.}(2021){Ashall}, {Lu}, {Hsiao}, {Hoeflich},
  {Phillips}, {Galbany}, {Burns}, {Contreras}, {Krisciunas}, {Morrell},
  {Stritzinger}, {Suntzeff}, {Taddia}, {Anais}, {Baron}, {Brown}, {Busta},
  {Campillay}, {Castell{\'o}n}, {Corco}, {Davis}, {Folatelli}, {F{\"o}rster},
  {Freedman}, {Gonzal{\'e}z}, {Hamuy}, {Holmbo}, {Kirshner}, {Kumar}, {Marion},
  {Mazzali}, {Morokuma}, {Nugent}, {Persson}, {Piro}, {Roth}, {Salgado},
  {Sand}, {Seron}, {Shahbandeh}, \& {Shappee}}]{2021ApJ...922..205A}
{Ashall}, C., {Lu}, J., {Hsiao}, E.~Y., {et~al.} 2021,
  \bibinfo{title}{{Carnegie Supernova Project: The First Homogeneous Sample of
  Super-Chandrasekhar-mass/2003fg-like Type Ia Supernovae},} \apj, 922, 205,
  \dodoi{10.3847/1538-4357/ac19ac}

\bibitem[{S. {Benetti} {et~al.}(2005){Benetti}, {Cappellaro}, {Mazzali},
  {Turatto}, {Altavilla}, {Bufano}, {Elias-Rosa}, {Kotak}, {Pignata}, {Salvo},
  \& {Stanishev}}]{benettietal05}
{Benetti}, S., {Cappellaro}, E., {Mazzali}, P.~A., {et~al.} 2005,
  \bibinfo{title}{{The Diversity of Type Ia Supernovae: Evidence for
  Systematics?},} \apj, 623, 1011, \dodoi{10.1086/428608}

\bibitem[{S. {Blondin} {et~al.}(2017){Blondin}, {Dessart}, {Hillier}, \&
  {Khokhlov}}]{blondinetal17}
{Blondin}, S., {Dessart}, L., {Hillier}, D.~J., \& {Khokhlov}, A.~M. 2017,
  \bibinfo{title}{{Evidence for sub-Chandrasekhar-mass progenitors of Type Ia
  supernovae at the faint end of the width-luminosity relation},} \mnras, 470,
  157, \dodoi{10.1093/mnras/stw2492}

\bibitem[{S. {Blondin} \& J.~L. {Tonry}(2007){Blondin} \&
  {Tonry}}]{blondintonry07}
{Blondin}, S., \& {Tonry}, J.~L. 2007, \bibinfo{title}{{Determining the Type,
  Redshift, and Age of a Supernova Spectrum},} \apj, 666, 1024,
  \dodoi{10.1086/520494}

\bibitem[{S. {Blondin} \& J.~L. {Tonry}(2011){Blondin} \&
  {Tonry}}]{blondin2011snid}
{Blondin}, S., \& {Tonry}, J.~L. 2011, {SNID: Supernova Identification},,
  Astrophysics Source Code Library, record ascl:1107.001

\bibitem[{T.~J. Boerner {et~al.}(2023)Boerner, Deems, Furlani, Knuth, \&
  Towns}]{boerner2023access}
Boerner, T.~J., Deems, S., Furlani, T.~R., Knuth, S.~L., \& Towns, J. 2023,
  \bibinfo{title}{ACCESS: Advancing Innovation: NSF's Advanced
  Cyberinfrastructure Coordination Ecosystem: Services \& Support,} in Practice
  and Experience in Advanced Research Computing 2023: Computing for the Common
  Good, PEARC '23 (New York, NY, USA: Association for Computing Machinery),
  173–176, \dodoi{10.1145/3569951.3597559}

\bibitem[{D. {Brout} {et~al.}(2022){Brout}, {Scolnic}, {Popovic}, {Riess},
  {Carr}, {Zuntz}, {Kessler}, {Davis}, {Hinton}, {Jones}, {Kenworthy},
  {Peterson}, {Said}, {Taylor}, {Ali}, {Armstrong}, {Charvu}, {Dwomoh},
  {Meldorf}, {Palmese}, {Qu}, {Rose}, {Sanchez}, {Stubbs}, {Vincenzi}, {Wood},
  {Brown}, {Chen}, {Chambers}, {Coulter}, {Dai}, {Dimitriadis}, {Filippenko},
  {Foley}, {Jha}, {Kelsey}, {Kirshner}, {Moller}, {Muir}, {Nadathur}, {Pan},
  {Rest}, {Rojas-Bravo}, {Sako}, {Siebert}, {Smith}, {Stahl}, \&
  {Wiseman}}]{Brout2022}
{Brout}, D., {Scolnic}, D., {Popovic}, B., {et~al.} 2022, \bibinfo{title}{{The
  Pantheon+ Analysis: Cosmological Constraints},} \apj, 938, 110,
  \dodoi{10.3847/1538-4357/ac8e04}

\bibitem[{A.~C. Calder {et~al.}(2007)Calder, Townsley, Seitenzahl, Peng,
  Messer, Vladimirova, Brown, Truran, \& Lamb}]{calderetal07}
Calder, A.~C., Townsley, D.~M., Seitenzahl, I.~R., {et~al.} 2007,
  \bibinfo{title}{{Capturing the Fire: Flame Energetics and Neutronizaton for
  Type Ia Supernova Simulations},} Astrophys. J., 656, 313,
  \dodoi{10.1086/510709}

\bibitem[{O. {Colin} {et~al.}(2000){Colin}, {Ducros}, {Veynante}, \&
  {Poinsot}}]{colin2000}
{Colin}, O., {Ducros}, F., {Veynante}, D., \& {Poinsot}, T. 2000,
  \bibinfo{title}{{A thickened flame model for large eddy simulations of
  turbulent premixed combustion},} Physics of Fluids, 12, 1843,
  \dodoi{10.1063/1.870436}

\bibitem[{S.~M. {Couch} {et~al.}(2013){Couch}, {Graziani}, \&
  {Flocke}}]{couch_2013}
{Couch}, S.~M., {Graziani}, C., \& {Flocke}, N. 2013, \bibinfo{title}{{An
  Improved Multipole Approximation for Self-gravity and Its Importance for
  Core-collapse Supernova Simulations},} \apj, 778, 181.
\newblock \url{http://dx.doi.org/10.1088/0004-637X/778/2/181}

\bibitem[{P. {Dave} {et~al.}(2017){Dave}, {Kashyap}, {Fisher}, {Timmes},
  {Townsley}, \& {Byrohl}}]{daveetal17}
{Dave}, P., {Kashyap}, R., {Fisher}, R., {et~al.} 2017,
  \bibinfo{title}{{Constraining the Single-degenerate Channel of Type Ia
  Supernovae with Stable Iron-group Elements in SNR 3C 397},} \apj, 841, 58,
  \dodoi{10.3847/1538-4357/aa7134}

\bibitem[{A. Dubey {et~al.}(2009)Dubey, Antypas, Ganapathy, Reid, Riley,
  Sheeler, Siegel, \& Weide}]{dubeyetal09}
Dubey, A., Antypas, K., Ganapathy, M.~K., {et~al.} 2009,
  \bibinfo{title}{Extensible Component Based Architecture for FLASH, A
  Massively Parallel, Multiphysics Simulation Code,} Parallel Computing, 35,
  512, \dodoi{http://dx.doi.org/10.1016/j.parco.2009.08.001}

\bibitem[{A. Dubey {et~al.}(2014)Dubey, Antypas, Calder, Daley, Fryxell,
  Gallagher, Lamb, Lee, Olson, Reid, Rich, Ricker, Riley, Rosner, Siegel,
  Taylor, Weide, Timmes, Vladimirova, \& ZuHone}]{Dubey14}
Dubey, A., Antypas, K., Calder, A.~C., {et~al.} 2014, \bibinfo{title}{Evolution
  of FLASH, a multi-physics scientific simulation code for high-performance
  computing,} The International Journal of High Performance Computing
  Applications, 28, 225, \dodoi{10.1177/1094342013505656}

\bibitem[{A.~V. {Filippenko}(1997){Filippenko}}]{filippenkoetal97}
{Filippenko}, A.~V. 1997, \bibinfo{title}{Optical Spectra of Supernovae,}
  \araa, 35, 309, \dodoi{10.1146/annurev.astro.35.1.309}

\bibitem[{A.~V. {Filippenko} {et~al.}(1992){Filippenko}, {Richmond},
  {Matheson}, {Shields}, {Burbidge}, {Cohen}, {Dickinson}, {Malkan}, {Nelson},
  {Pietz}, {Schlegel}, {Schmeer}, {Spinrad}, {Steidel}, {Tran}, \&
  {Wren}}]{filippenkoetal92}
{Filippenko}, A.~V., {Richmond}, M.~W., {Matheson}, T., {et~al.} 1992,
  \bibinfo{title}{{The peculiar Type IA SN 1991T - Detonation of a white
  dwarf?},} \apjl, 384, L15, \dodoi{10.1086/186252}

\bibitem[{R.~J. {Foley} {et~al.}(2009){Foley}, {Matheson}, {Blondin},
  {Chornock}, {Silverman}, {Challis}, {Clocchiatti}, {Filippenko}, {Kirshner},
  {Leibundgut}, {Sollerman}, {Spyromilio}, {Tonry}, {Davis}, {Garnavich},
  {Jha}, {Krisciunas}, {Li}, {Pignata}, {Rest}, {Riess}, {Schmidt}, {Smith},
  {Stubbs}, {Tucker}, \& {Wood-Vasey}}]{foleyetal09}
{Foley}, R.~J., {Matheson}, T., {Blondin}, S., {et~al.} 2009,
  \bibinfo{title}{{Spectroscopy of High-Redshift Supernovae from the Essence
  Project: The First Four Years},} \aj, 137, 3731,
  \dodoi{10.1088/0004-6256/137/4/3731}

\bibitem[{B. {Fryxell} {et~al.}(2000){Fryxell}, {Olson}, {Ricker}, {Timmes},
  {Zingale}, {Lamb}, {MacNeice}, {Rosner}, {Truran}, \&
  {Tufo}}]{fryxell2000flash}
{Fryxell}, B., {Olson}, K., {Ricker}, P., {et~al.} 2000,
  \bibinfo{title}{{FLASH: An Adaptive Mesh Hydrodynamics Code for Modeling
  Astrophysical Thermonuclear Flashes},} \apjs, 131, 273,
  \dodoi{10.1086/317361}

\bibitem[{V.~N. {Gamezo} {et~al.}(2004){Gamezo}, {Khokhlov}, \&
  {Oran}}]{gamezoetal04}
{Gamezo}, V.~N., {Khokhlov}, A.~M., \& {Oran}, E.~S. 2004,
  \bibinfo{title}{{Deflagrations and Detonations in Thermonuclear Supernovae},}
  Physical Review Letters, 92, 211102, \dodoi{10.1103/PhysRevLett.92.211102}

\bibitem[{V.~N. {Gamezo} {et~al.}(2005){Gamezo}, {Khokhlov}, \&
  {Oran}}]{gamezoetal05}
{Gamezo}, V.~N., {Khokhlov}, A.~M., \& {Oran}, E.~S. 2005,
  \bibinfo{title}{{Three-dimensional Delayed-Detonation Model of Type Ia
  Supernovae},} \apj, 623, 337, \dodoi{10.1086/428767}

\bibitem[{S. Gronow {et~al.}(2021)Gronow, Collins, Sim, \&
  Röpke}]{gronowetal2021}
Gronow, S., Collins, C.~E., Sim, S.~A., \& Röpke, F.~K. 2021,
  \bibinfo{title}{Double detonations of sub-MCh CO white dwarfs: variations in
  Type Ia supernovae due to different core and He shell masses,} \aap, 649,
  A155, \dodoi{10.1051/0004-6361/202039954}

\bibitem[{P.~E. Hamlington {et~al.}(2011)Hamlington, Poludnenko, \&
  Oran}]{Hamlington2011}
Hamlington, P.~E., Poludnenko, A.~Y., \& Oran, E.~S. 2011,
  \bibinfo{title}{Interactions between turbulence and flames in premixed
  reacting flows,} Physics of Fluids, 23, \dodoi{10.1063/1.3671736}

\bibitem[{P.~E. Hamlington {et~al.}(2012)Hamlington, Poludnenko, \&
  Oran}]{Hamlington2012}
Hamlington, P.~E., Poludnenko, A.~Y., \& Oran, E.~S. 2012,
  \bibinfo{title}{Intermittency in premixed turbulent reacting flows,} Physics
  of Fluids, 24, \dodoi{10.1063/1.4729615}

\bibitem[{F. {Hoyle} \& W.~A. {Fowler}(1960){Hoyle} \&
  {Fowler}}]{hoylefowler60}
{Hoyle}, F., \& {Fowler}, W.~A. 1960, \bibinfo{title}{{Nucleosynthesis in
  Supernovae.},} \apj, 132, 565, \dodoi{10.1086/146963}

\bibitem[{J.~D. Hunter(2007)Hunter}]{hunter07}
Hunter, J.~D. 2007, \bibinfo{title}{Matplotlib: A 2D Graphics Environment,}
  Computing in Science Engineering, 9, 90, \dodoi{10.1109/MCSE.2007.55}

\bibitem[{I. Iben \& A.~V. Tutukov(1984)Iben \& Tutukov}]{iben1984supernovae}
Iben, Jr, I., \& Tutukov, A.~V. 1984, \bibinfo{title}{Supernovae of type I as
  end products of the evolution of binaries with components of moderate initial
  mass (M not greater than about 9 solar masses),} The Astrophysical Journal
  Supplement Series, 54, 335

\bibitem[{K. Iwamoto {et~al.}(1999)Iwamoto, Brachwitz, Nomoto, Kishimoto,
  Umeda, Hix, \& Thielemann}]{iwamoto1999nucleosynthesis}
Iwamoto, K., Brachwitz, F., Nomoto, K., {et~al.} 1999,
  \bibinfo{title}{Nucleosynthesis in Chandrasekhar mass models for type Ia
  supernovae and constraints on progenitor systems and burning-front
  propagation,} The Astrophysical Journal Supplement Series, 125, 439

\bibitem[{A.~P. {Jackson} {et~al.}(2010){Jackson}, {Calder}, {Townsley},
  {Chamulak}, {Brown}, \& {Timmes}}]{jacksonetal2010}
{Jackson}, A.~P., {Calder}, A.~C., {Townsley}, D.~M., {et~al.} 2010,
  \bibinfo{title}{{Evaluating Systematic Dependencies of Type Ia Supernovae:
  The Influence of Deflagration to Detonation Density},} \apj, 720, 99,
  \dodoi{10.1088/0004-637X/720/1/99}

\bibitem[{D.~O. {Jones} {et~al.}(2018){Jones}, {Riess}, {Scolnic}, {Pan},
  {Johnson}, {Coulter}, {Dettman}, {Foley}, {Foley}, {Huber}, {Jha},
  {Kilpatrick}, {Kirshner}, {Rest}, {Rojas-Bravo}, \& {Siebert}}]{Jones2018}
{Jones}, D.~O., {Riess}, A.~G., {Scolnic}, D.~M., {et~al.} 2018,
  \bibinfo{title}{{Should Type Ia Supernova Distances Be Corrected for Their
  Local Environments?},} \apj, 867, 108, \dodoi{10.3847/1538-4357/aae2b9}

\bibitem[{G.~C. {Jordan} {et~al.}(2012){Jordan}, {Perets}, {Fisher}, \& {van
  Rossum}}]{jordanetal12b}
{Jordan}, IV, G.~C., {Perets}, H.~B., {Fisher}, R.~T., \& {van Rossum}, D.~R.
  2012, \bibinfo{title}{{Failed-detonation Supernovae: Subluminous Low-velocity
  Ia Supernovae and their Kicked Remnant White Dwarfs with Iron-rich Cores},}
  \apjl, 761, L23, \dodoi{10.1088/2041-8205/761/2/L23}

\bibitem[{A.~M. {Khokhlov}(1991){Khokhlov}}]{khokhlov91}
{Khokhlov}, A.~M. 1991, \bibinfo{title}{{Delayed detonation model for type IA
  supernovae},} \aap, 245, 114

\bibitem[{A.~M. {Khokhlov} {et~al.}(1997){Khokhlov}, {Oran}, \&
  {Wheeler}}]{khokhlovetal97}
{Khokhlov}, A.~M., {Oran}, E.~S., \& {Wheeler}, J.~C. 1997,
  \bibinfo{title}{{Deflagration-to-Detonation Transition in Thermonuclear
  Supernovae},} \apj, 478, 678, \dodoi{10.1086/303815}

\bibitem[{C. {Kobayashi} {et~al.}(2006){Kobayashi}, {Umeda}, {Nomoto},
  {Tominaga}, \& {Ohkubo}}]{kobayashietal06}
{Kobayashi}, C., {Umeda}, H., {Nomoto}, K., {Tominaga}, N., \& {Ohkubo}, T.
  2006, \bibinfo{title}{{Galactic Chemical Evolution: Carbon through Zinc},}
  \apj, 653, 1145, \dodoi{10.1086/508914}

\bibitem[{B.~K. {Krueger} {et~al.}(2012){Krueger}, {Jackson}, {Calder},
  {Townsley}, {Brown}, \& {Timmes}}]{kruegeretal12}
{Krueger}, B.~K., {Jackson}, A.~P., {Calder}, A.~C., {et~al.} 2012,
  \bibinfo{title}{{Evaluating Systematic Dependencies of Type Ia Supernovae:
  The Influence of Central Density},} \apj, 757, 175,
  \dodoi{10.1088/0004-637X/757/2/175}

\bibitem[{L.~D. {Landau} \& E.~M. {Lifshitz}(1959){Landau} \&
  {Lifshitz}}]{landaulifshitz59}
{Landau}, L.~D., \& {Lifshitz}, E.~M. 1959, {Fluid mechanics} (Pergamon Press)

\bibitem[{P. {Lesaffre} {et~al.}(2006){Lesaffre}, {Han}, {Tout},
  {Podsiadlowski}, \& {Martin}}]{lesaffreetal06}
{Lesaffre}, P., {Han}, Z., {Tout}, C.~A., {Podsiadlowski}, P., \& {Martin},
  R.~G. 2006, \bibinfo{title}{{The C flash and the ignition conditions of Type
  Ia supernovae},} \mnras, 368, 187, \dodoi{10.1111/j.1365-2966.2006.10068.x}

\bibitem[{S.-C. {Leung} \& K. {Nomoto}(2018){Leung} \&
  {Nomoto}}]{leungnomoto18}
{Leung}, S.-C., \& {Nomoto}, K. 2018, \bibinfo{title}{{Explosive
  Nucleosynthesis in Near-Chandrasekhar-mass White Dwarf Models for Type Ia
  Supernovae: Dependence on Model Parameters},} \apj, 861, 143,
  \dodoi{10.3847/1538-4357/aac2df}

\bibitem[{K. {Lodders} {et~al.}(2025){Lodders}, {Bergemann}, \&
  {Palme}}]{lodders_2025}
{Lodders}, K., {Bergemann}, M., \& {Palme}, H. 2025, \bibinfo{title}{Solar
  System Elemental Abundances from the Solar Photosphere and CI-Chondrites,}
  Space Sci Rev, 221, 23, \dodoi{https://doi.org/10.1007/s11214-025-01146-w}

\bibitem[{D. Maoz {et~al.}(2014)Maoz, Mannucci, \&
  Nelemans}]{maoz2014observational}
Maoz, D., Mannucci, F., \& Nelemans, G. 2014, \bibinfo{title}{Observational
  clues to the progenitors of type Ia supernovae,} Annual Review of Astronomy
  and Astrophysics, 52, 107

\bibitem[{D. {Maoz} {et~al.}(2014){Maoz}, {Mannucci}, \& {Nelemans}}]{Maoz2014}
{Maoz}, D., {Mannucci}, F., \& {Nelemans}, G. 2014,
  \bibinfo{title}{{Observational Clues to the Progenitors of Type Ia
  Supernovae},} \araa, 52, 107, \dodoi{10.1146/annurev-astro-082812-141031}

\bibitem[{E. {Mueller} \& W.~D. {Arnett}(1982){Mueller} \&
  {Arnett}}]{muellerarnett82}
{Mueller}, E., \& {Arnett}, W.~D. 1982, \bibinfo{title}{{Numerical studies of
  nonspherical carbon combustion models},} \apjl, 261, L109,
  \dodoi{10.1086/183898}

\bibitem[{J.~C. {Niemeyer} \& S.~E. {Woosley}(1997){Niemeyer} \&
  {Woosley}}]{niemeyerwoosley97}
{Niemeyer}, J.~C., \& {Woosley}, S.~E. 1997, \bibinfo{title}{{The Thermonuclear
  Explosion of Chandrasekhar Mass White Dwarfs},} \apj, 475, 740,
  \dodoi{10.1086/303544}

\bibitem[{K. {Nomoto}(1982){Nomoto}}]{nomoto1982accreting}
{Nomoto}, K. 1982, \bibinfo{title}{{Accreting white dwarf models for type 1
  supernovae. II - Off-center detonation supernovae},} \apj, 257, 780,
  \dodoi{10.1086/160031}

\bibitem[{K. {Nomoto} {et~al.}(1997){Nomoto}, {Iwamoto}, \&
  {Kishimoto}}]{nomoto1997Sci...276.1378N}
{Nomoto}, K., {Iwamoto}, K., \& {Kishimoto}, N. 1997, \bibinfo{title}{{Type Ia
  supernovae: their origin and possible applications in cosmology.},} Science,
  276, 1378, \dodoi{10.1126/science.276.5317.1378}

\bibitem[{K. {Nomoto} {et~al.}(1984{\natexlab{a}}){Nomoto}, Thielemann, \&
  Yokoi}]{nomoto1984accreting}
{Nomoto}, K., Thielemann, F.-K., \& Yokoi, K. 1984{\natexlab{a}},
  \bibinfo{title}{Accreting white dwarf models of Type I supernovae. III-Carbon
  deflagration supernovae,} The Astrophysical Journal, 286, 644

\bibitem[{K. {Nomoto} {et~al.}(1984{\natexlab{b}}){Nomoto}, {Thielemann}, \&
  {Yokoi}}]{nomotoetal84}
{Nomoto}, K., {Thielemann}, F.-K., \& {Yokoi}, K. 1984{\natexlab{b}},
  \bibinfo{title}{{Accreting white dwarf models of Type I supernovae. III -
  Carbon deflagration supernovae},} \apj, 286, 644, \dodoi{10.1086/162639}

\bibitem[{A. {Nonaka} {et~al.}(2012){Nonaka}, {Aspden}, {Zingale}, {Almgren},
  {Bell}, \& {Woosley}}]{nonakaetal12}
{Nonaka}, A., {Aspden}, A.~J., {Zingale}, M., {et~al.} 2012,
  \bibinfo{title}{{High-resolution Simulations of Convection Preceding Ignition
  in Type Ia Supernovae Using Adaptive Mesh Refinement},} \apj, 745, 73,
  \dodoi{10.1088/0004-637X/745/1/73}

\bibitem[{S.~T. {Ohlmann} {et~al.}(2014){Ohlmann}, {Kromer}, {Fink}, {Pakmor},
  {Seitenzahl}, {Sim}, \& {R{\"o}pke}}]{ohlmannetal14}
{Ohlmann}, S.~T., {Kromer}, M., {Fink}, M., {et~al.} 2014, \bibinfo{title}{{The
  white dwarf's carbon fraction as a secondary parameter of Type Ia
  supernovae},} \aap, 572, A57, \dodoi{10.1051/0004-6361/201423924}

\bibitem[{Y. {Ohshiro} {et~al.}(2021){Ohshiro}, {Yamaguchi}, {Leung}, {Nomoto},
  {Sato}, {Tanaka}, {Okon}, {Fisher}, {Petre}, \& {Williams}}]{oshiroetal21}
{Ohshiro}, Y., {Yamaguchi}, H., {Leung}, S.-C., {et~al.} 2021,
  \bibinfo{title}{{Discovery of a Highly Neutronized Ejecta Clump in the Type
  Ia Supernova Remnant 3C 397},} \apjl, 913, L34,
  \dodoi{10.3847/2041-8213/abff5b}

\bibitem[{F. Perez \& B.~E. Granger(2007)Perez \& Granger}]{perezgranger07}
Perez, F., \& Granger, B.~E. 2007, \bibinfo{title}{IPython: A System for
  Interactive Scientific Computing,} Computing in Science Engineering, 9, 21,
  \dodoi{10.1109/MCSE.2007.53}

\bibitem[{S. Perlmutter {et~al.}(1999)Perlmutter, Aldering, Goldhaber, Knop,
  Nugent, Castro, Deustua, Fabbro, Goobar, Groom, \& et~al.}]{perlmutter_1999}
Perlmutter, S., Aldering, G., Goldhaber, G., {et~al.} 1999,
  \bibinfo{title}{{Measurements of $\Omega$ and $\Lambda$ from 42 High Redshift
  Supernovae},} The Astrophysical Journal, 517, 565{\textendash}586,
  \dodoi{10.1086/307221}

\bibitem[{A.~Y. {Poludnenko} {et~al.}(2019){Poludnenko}, {Chambers}, {Ahmed},
  {Gamezo}, \& {Taylor}}]{poludnenkoetal19}
{Poludnenko}, A.~Y., {Chambers}, J., {Ahmed}, K., {Gamezo}, V.~N., \& {Taylor},
  B.~D. 2019, \bibinfo{title}{{A unified mechanism for unconfined
  deflagration-to-detonation transition in terrestrial chemical systems and
  type Ia supernovae},} Science, 366, aau7365, \dodoi{10.1126/science.aau7365}

\bibitem[{A.~Y. {Poludnenko} {et~al.}(2011){Poludnenko}, {Gardiner}, \&
  {Oran}}]{poludnenkoetal11}
{Poludnenko}, A.~Y., {Gardiner}, T.~A., \& {Oran}, E.~S. 2011,
  \bibinfo{title}{{Spontaneous Transition of Turbulent Flames to Detonations in
  Unconfined Media},} Physical Review Letters, 107, 054501,
  \dodoi{10.1103/PhysRevLett.107.054501}

\bibitem[{R. Raddi {et~al.}(2018)Raddi, Hollands, Gänsicke, Townsley, Hermes,
  Gentile Fusillo, \& Koester}]{raddietal18}
Raddi, R., Hollands, M.~A., Gänsicke, B.~T., {et~al.} 2018,
  \bibinfo{title}{{Anatomy of the hyper-runaway star LP 40–365 with Gaia},}
  Monthly Notices of the Royal Astronomical Society: Letters, 479, L96,
  \dodoi{10.1093/mnrasl/sly103}

\bibitem[{A.~G. Riess {et~al.}(1998)Riess, Filippenko, Challis, Clocchiatti,
  Diercks, Garnavich, Gilliland, Hogan, Jha, Kirshner, \& et~al.}]{Riess_1998}
Riess, A.~G., Filippenko, A.~V., Challis, P., {et~al.} 1998,
  \bibinfo{title}{{Observational Evidence from Supernovae for an Accelerating
  Universe and a Cosmological Constant},} The Astronomical Journal, 116,
  1009{\textendash}1038, \dodoi{10.1086/300499}

\bibitem[{A.~G. {Riess} {et~al.}(1998){Riess}, {Filippenko}, {Challis},
  {Clocchiatti}, {Diercks}, {Garnavich}, {Gilliland}, {Hogan}, {Jha},
  {Kirshner}, {Leibundgut}, {Phillips}, {Reiss}, {Schmidt}, {Schommer},
  {Smith}, {Spyromilio}, {Stubbs}, {Suntzeff}, \&
  {Tonry}}]{riess..1998AJ....116.1009R}
{Riess}, A.~G., {Filippenko}, A.~V., {Challis}, P., {et~al.} 1998,
  \bibinfo{title}{{Observational Evidence from Supernovae for an Accelerating
  Universe and a Cosmological Constant},} \aj, 116, 1009,
  \dodoi{10.1086/300499}

\bibitem[{M. {Rigault} {et~al.}(2020){Rigault}, {Brinnel}, {Aldering},
  {Antilogus}, {Aragon}, {Bailey}, {Baltay}, {Barbary}, {Bongard}, {Boone},
  {Buton}, {Childress}, {Chotard}, {Copin}, {Dixon}, {Fagrelius}, {Feindt},
  {Fouchez}, {Gangler}, {Hayden}, {Hillebrandt}, {Howell}, {Kim}, {Kowalski},
  {Kuesters}, {Leget}, {Lombardo}, {Lin}, {Nordin}, {Pain}, {Pecontal},
  {Pereira}, {Perlmutter}, {Rabinowitz}, {Runge}, {Rubin}, {Saunders},
  {Smadja}, {Sofiatti}, {Suzuki}, {Taubenberger}, {Tao}, \&
  {Thomas}}]{Rigault2020}
{Rigault}, M., {Brinnel}, V., {Aldering}, G., {et~al.} 2020,
  \bibinfo{title}{{Strong dependence of Type Ia supernova standardization on
  the local specific star formation rate},} \aap, 644, A176,
  \dodoi{10.1051/0004-6361/201730404}

\bibitem[{F. R{\"o}pke {et~al.}(2012)R{\"o}pke, Kromer, Seitenzahl, Pakmor,
  Sim, Taubenberger, Ciaraldi-Schoolmann, Hillebrandt, Aldering, Antilogus,
  {et~al.}}]{ropke2012constraining}
R{\"o}pke, F., Kromer, M., Seitenzahl, I., {et~al.} 2012,
  \bibinfo{title}{Constraining type Ia supernova models: SN 2011fe as a test
  case,} The Astrophysical Journal Letters, 750, L19

\bibitem[{F.~K. {R{\"o}pke} {et~al.}(2006){R{\"o}pke}, {Gieseler}, {Reinecke},
  {Travaglio}, \& {Hillebrandt}}]{ropkeetal06}
{R{\"o}pke}, F.~K., {Gieseler}, M., {Reinecke}, M., {Travaglio}, C., \&
  {Hillebrandt}, W. 2006, \bibinfo{title}{{Type Ia supernova diversity in
  three-dimensional models},} \aap, 453, 203,
  \dodoi{10.1051/0004-6361:20053430}

\bibitem[{F.~K. {R{\"o}pke} {et~al.}(2007{\natexlab{a}}){R{\"o}pke},
  {Hillebrandt}, {Schmidt}, {Niemeyer}, {Blinnikov}, \&
  {Mazzali}}]{ropkeetal07}
{R{\"o}pke}, F.~K., {Hillebrandt}, W., {Schmidt}, W., {et~al.}
  2007{\natexlab{a}}, \bibinfo{title}{{A Three-Dimensional Deflagration Model
  for Type Ia Supernovae Compared with Observations},} \apj, 668, 1132,
  \dodoi{10.1086/521347}

\bibitem[{F.~K. {R{\"o}pke} {et~al.}(2007{\natexlab{b}}){R{\"o}pke}, {Woosley},
  \& {Hillebrandt}}]{roepkeetal07}
{R{\"o}pke}, F.~K., {Woosley}, S.~E., \& {Hillebrandt}, W. 2007{\natexlab{b}},
  \bibinfo{title}{{Off-Center Ignition in Type Ia Supernovae. I. Initial
  Evolution and Implications for Delayed Detonation},} \apj, 660, 1344,
  \dodoi{10.1086/512769}

\bibitem[{P. {Ruiz-Lapuente} {et~al.}(1992){Ruiz-Lapuente}, {Cappellaro},
  {Turatto}, {Gouiffes}, {Danziger}, {della Valle}, \&
  {Lucy}}]{ruizlapuenteetal92}
{Ruiz-Lapuente}, P., {Cappellaro}, E., {Turatto}, M., {et~al.} 1992,
  \bibinfo{title}{{Modeling the iron-dominated spectra of the type IA supernova
  SN 1991T at premaximum},} \apjl, 387, L33, \dodoi{10.1086/186299}

\bibitem[{B.~P. Schmidt {et~al.}(1998)Schmidt, Suntzeff, Phillips, Schommer,
  Clocchiatti, Kirshner, Garnavich, Challis, Leibundgut, Spyromilio, \&
  et~al.}]{schmidt_1998}
Schmidt, B.~P., Suntzeff, N.~B., Phillips, M.~M., {et~al.} 1998,
  \bibinfo{title}{{The High{}Z Supernova Search: Measuring Cosmic Deceleration
  and Global Curvature of the Universe Using Type Ia Supernovae},} The
  Astrophysical Journal, 507, 46{\textendash}63, \dodoi{10.1086/306308}

\bibitem[{I.~R. {Seitenzahl} {et~al.}(2013{\natexlab{a}}){Seitenzahl},
  {Cescutti}, {R{\"o}pke}, {Ruiter}, \& {Pakmor}}]{seitenzahletal13}
{Seitenzahl}, I.~R., {Cescutti}, G., {R{\"o}pke}, F.~K., {Ruiter}, A.~J., \&
  {Pakmor}, R. 2013{\natexlab{a}}, \bibinfo{title}{{Solar abundance of
  manganese: a case for near Chandrasekhar-mass Type Ia supernova
  progenitors},} \aap, 559, L5, \dodoi{10.1051/0004-6361/201322599}

\bibitem[{I.~R. {Seitenzahl} \& D.~M. {Townsley}(2017){Seitenzahl} \&
  {Townsley}}]{seitenzahltownsley17}
{Seitenzahl}, I.~R., \& {Townsley}, D.~M. 2017,
  \bibinfo{title}{{Nucleosynthesis in Thermonuclear Supernovae},} in Handbook
  of Supernovae, ed. A.~W. {Alsabti} \& P.~{Murdin}, 1955,
  \dodoi{10.1007/978-3-319-21846-5_87}

\bibitem[{I.~R. {Seitenzahl} {et~al.}(2013{\natexlab{b}}){Seitenzahl},
  {Ciaraldi-Schoolmann}, {R{\"o}pke}, {Fink}, {Hillebrandt}, {Kromer},
  {Pakmor}, {Ruiter}, {Sim}, \& {Taubenberger}}]{seitenzahletal13b}
{Seitenzahl}, I.~R., {Ciaraldi-Schoolmann}, F., {R{\"o}pke}, F.~K., {et~al.}
  2013{\natexlab{b}}, \bibinfo{title}{{Three-dimensional delayed-detonation
  models with nucleosynthesis for Type Ia supernovae},} Monthly Notices of the
  Royal Astronomical Society, 429, 1156, \dodoi{10.1093/mnras/sts402}

\bibitem[{I.~R. {Seitenzahl} {et~al.}(2015){Seitenzahl}, {Summa}, {Krau{\ss}},
  {Sim}, {Diehl}, {Els{\"a}sser}, {Fink}, {Hillebrandt}, {Kromer}, {Maeda},
  {Mannheim}, {Pakmor}, {R{\"o}pke}, {Ruiter}, \& {Wilms}}]{seitenzahletal15}
{Seitenzahl}, I.~R., {Summa}, A., {Krau{\ss}}, F., {et~al.} 2015,
  \bibinfo{title}{{5.9-keV Mn K-shell X-ray luminosity from the decay of
  $^{55}$Fe in Type Ia supernova models},} \mnras, 447, 1484,
  \dodoi{10.1093/mnras/stu2537}

\bibitem[{K.~J. Shen {et~al.}(2018)Shen, Boubert, G{\"a}nsicke, Jha, Andrews,
  Chomiuk, Foley, Fraser, Gromadzki, Guillochon, {et~al.}}]{shen2018three}
Shen, K.~J., Boubert, D., G{\"a}nsicke, B.~T., {et~al.} 2018,
  \bibinfo{title}{Three hypervelocity white dwarfs in Gaia DR2: Evidence for
  dynamically driven double-degenerate double-detonation type Ia supernovae,}
  The Astrophysical Journal, 865, 15

\bibitem[{J.~M. {Silverman} {et~al.}(2012){Silverman}, {Foley}, {Filippenko},
  {Ganeshalingam}, {Barth}, {Chornock}, {Griffith}, {Kong}, {Lee}, {Leonard},
  {Matheson}, {Miller}, {Steele}, {Barris}, {Bloom}, {Cobb}, {Coil},
  {Desroches}, {Gates}, {Ho}, {Jha}, {Kandrashoff}, {Li}, {Mandel}, {Modjaz},
  {Moore}, {Mostardi}, {Papenkova}, {Park}, {Perley}, {Poznanski}, {Reuter},
  {Scala}, {Serduke}, {Shields}, {Swift}, {Tonry}, {Van Dyk}, {Wang}, \&
  {Wong}}]{silvermanetal12_bsnip}
{Silverman}, J.~M., {Foley}, R.~J., {Filippenko}, A.~V., {et~al.} 2012,
  \bibinfo{title}{{Berkeley Supernova Ia Program -- I. Observations, data
  reduction and spectroscopic sample of 582 low-redshift Type Ia supernovae},}
  \mnras, 425, 1789, \dodoi{10.1111/j.1365-2966.2012.21270.x}

\bibitem[{S.~A. {Sim} {et~al.}(2013){Sim}, {Seitenzahl}, {Kromer},
  {Ciaraldi-Schoolmann}, {R{\"o}pke}, {Fink}, {Hillebrandt}, {Pakmor},
  {Ruiter}, \& {Taubenberger}}]{simetal13}
{Sim}, S.~A., {Seitenzahl}, I.~R., {Kromer}, M., {et~al.} 2013,
  \bibinfo{title}{{Synthetic light curves and spectra for three-dimensional
  delayed-detonation models of Type Ia supernovae},} \mnras, 436, 333,
  \dodoi{10.1093/mnras/stt1574}

\bibitem[{M. {Stritzinger} {et~al.}(2006){Stritzinger}, {Mazzali}, {Sollerman},
  \& {Benetti}}]{stritzingeretal06}
{Stritzinger}, M., {Mazzali}, P.~A., {Sollerman}, J., \& {Benetti}, S. 2006,
  \bibinfo{title}{{Consistent estimates of $^{56}$Ni yields for type Ia
  supernovae},} \aap, 460, 793, \dodoi{10.1051/0004-6361:20065514}

\bibitem[{L.-S. The {et~al.}(2006)The, Clayton, Diehl, Hartmann, Iyudin,
  Leising, Meyer, Motizuki, \& Schönfelder}]{The2006}
The, L.-S., Clayton, D.~D., Diehl, R., {et~al.} 2006, \bibinfo{title}{Are
  $\mathsf{^{44} }$Ti-producing supernovae exceptional?} Astronomy \&
  Astrophysics, 450, 1037–1050, \dodoi{10.1051/0004-6361:20054626}

\bibitem[{F.~X. Timmes(1999)Timmes}]{timmes_1999}
Timmes, F.~X. 1999, \bibinfo{title}{{Integration of Nuclear Reaction Networks
  for Stellar Hydrodynamics},} The Astrophysical Journal Supplement Series,
  124, 241{\textendash}263, \dodoi{10.1086/313257}

\bibitem[{F.~X. {Timmes} \& S.~E. {Woosley}(1992){Timmes} \&
  {Woosley}}]{timmeswoosley92}
{Timmes}, F.~X., \& {Woosley}, S.~E. 1992, \bibinfo{title}{{The conductive
  propagation of nuclear flames. I - Degenerate C + O and O + NE + MG white
  dwarfs},} \apj, 396, 649, \dodoi{10.1086/171746}

\bibitem[{J. {Tonry} \& M. {Davis}(1979){Tonry} \& {Davis}}]{tonry_1979}
{Tonry}, J., \& {Davis}, M. 1979, \bibinfo{title}{{A survey of galaxy
  redshifts. I. Data reduction techniques.},} The Astronomical Journal, 84,
  1511, \dodoi{10.1086/112569}

\bibitem[{D.~M. {Townsley} {et~al.}(2007){Townsley}, {Calder}, {Asida},
  {Seitenzahl}, {Peng}, {Vladimirova}, {Lamb}, \& {Truran}}]{townsleyetal07}
{Townsley}, D.~M., {Calder}, A.~C., {Asida}, S.~M., {et~al.} 2007,
  \bibinfo{title}{{Flame Evolution During Type Ia Supernovae and the
  Deflagration Phase in the Gravitationally Confined Detonation Scenario},}
  \apj, 668, 1118, \dodoi{10.1086/521013}

\bibitem[{D.~M. {Townsley} {et~al.}(2009){Townsley}, {Jackson}, {Calder},
  {Chamulak}, {Brown}, \& {Timmes}}]{townsleyetal09}
{Townsley}, D.~M., {Jackson}, A.~P., {Calder}, A.~C., {et~al.} 2009,
  \bibinfo{title}{{Evaluating Systematic Dependencies of Type Ia Supernovae:
  The Influence of Progenitor $^{22}$Ne Content on Dynamics},} \apj, 701, 1582,
  \dodoi{10.1088/0004-637X/701/2/1582}

\bibitem[{D.~M. {Townsley} {et~al.}(2016){Townsley}, {Miles}, {Timmes},
  {Calder}, \& {Brown}}]{townsleyetal16}
{Townsley}, D.~M., {Miles}, B.~J., {Timmes}, F.~X., {Calder}, A.~C., \&
  {Brown}, E.~F. 2016, \bibinfo{title}{{A Tracer Method for Computing Type Ia
  Supernova Yields: Burning Model Calibration, Reconstruction of Thickened
  Flames, and Verification for Planar Detonations},} \apjs, 225, 3,
  \dodoi{10.3847/0067-0049/225/1/3}

\bibitem[{M.~J. {Turk} {et~al.}(2011){Turk}, {Smith}, {Oishi}, {Skory},
  {Skillman}, {Abel}, \& {Norman}}]{turk_2011}
{Turk}, M.~J., {Smith}, B.~D., {Oishi}, J.~S., {et~al.} 2011,
  \bibinfo{title}{{yt: A Multi-code Analysis Toolkit for Astrophysical
  Simulation Data},} \apjs, 192, 9.
\newblock \url{http://dx.doi.org/10.1088/0067-0049/192/1/9}

\bibitem[{S. van~der Walt {et~al.}(2011)van~der Walt, Colbert, \&
  Varoquaux}]{vanderwaltetal11}
van~der Walt, S., Colbert, S.~C., \& Varoquaux, G. 2011, \bibinfo{title}{The
  NumPy Array: A Structure for Efficient Numerical Computation,} Computing in
  Science Engineering, 13, 22, \dodoi{10.1109/MCSE.2011.37}

\bibitem[{D.~R. van Rossum {et~al.}(2016)van Rossum, Kashyap, Fisher,
  Wollaeger, García-Berro, Aznar-Siguán, Ji, \&
  Lorén-Aguilar}]{van_Rossum_2016}
van Rossum, D.~R., Kashyap, R., Fisher, R., {et~al.} 2016,
  \bibinfo{title}{LIGHT CURVES AND SPECTRA FROM A THERMONUCLEAR EXPLOSION OF A
  WHITE DWARF MERGER,} The Astrophysical Journal, 827, 128,
  \dodoi{10.3847/0004-637X/827/2/128}

\bibitem[{G. van Rossum \& J. de~Boer(1991)van Rossum \&
  de~Boer}]{vanrossumdeboer91}
van Rossum, G., \& de~Boer, J. 1991, \bibinfo{title}{Interactively testing
  remote servers using the Python programming language,} CWI Quarterly, 4, 283

\bibitem[{P. Virtanen {et~al.}(2020)Virtanen, Gommers, Oliphant, Haberland,
  Reddy, Cournapeau, Burovski, Peterson, Weckesser, Bright, {van der Walt},
  Brett, Wilson, Millman, Mayorov, Nelson, Jones, Kern, Larson, Carey, Polat,
  Feng, Moore, {VanderPlas}, Laxalde, Perktold, Cimrman, Henriksen, Quintero,
  Harris, Archibald, Ribeiro, Pedregosa, {van Mulbregt}, \& {SciPy 1.0
  Contributors}}]{2020SciPy-NMeth}
Virtanen, P., Gommers, R., Oliphant, T.~E., {et~al.} 2020,
  \bibinfo{title}{{{SciPy} 1.0: Fundamental Algorithms for Scientific Computing
  in Python},} Nature Methods, 17, 261, \dodoi{10.1038/s41592-019-0686-2}

\bibitem[{R.~F. {Webbink}(1984){Webbink}}]{webbink84}
{Webbink}, R.~F. 1984, \bibinfo{title}{{Double white dwarfs as progenitors of R
  Coronae Borealis stars and Type I supernovae},} \apj, 277, 355,
  \dodoi{10.1086/161701}

\bibitem[{J. Whelan \& I. Iben~Jr(1973)Whelan \& Iben~Jr}]{whelan1973binaries}
Whelan, J., \& Iben~Jr, I. 1973, \bibinfo{title}{Binaries and supernovae of
  type I,} The Astrophysical Journal, 186, 1007

\bibitem[{R.~T. Wollaeger \& D.~R. van Rossum(2014)Wollaeger \& van
  Rossum}]{wollaeger_2014}
Wollaeger, R.~T., \& van Rossum, D.~R. 2014, \bibinfo{title}{{Radiation
  Transport for Explosive Outflows: Opacity Regrouping},} The Astrophysical
  Journal Supplement Series, 214, 21, \dodoi{10.1088/0067-0049/214/2/28}

\bibitem[{R.~T. Wollaeger {et~al.}(2013)Wollaeger, van Rossum, Graziani, Couch,
  Jordan, Lamb, \& Moses}]{wollaeger_2013}
Wollaeger, R.~T., van Rossum, D.~R., Graziani, C., {et~al.} 2013,
  \bibinfo{title}{{Radiation Transport for Explosive Outflows: A Multigroup
  Hybrid Monte Carlo Method},} The Astrophysical Journal Supplement Series,
  209, 21, \dodoi{10.1088/0067-0049/209/2/36}

\bibitem[{S.~E. {Woosley}(2007){Woosley}}]{woosley07}
{Woosley}, S.~E. 2007, \bibinfo{title}{{Type Ia Supernovae: Burning and
  Detonation in the Distributed Regime},} \apj, 668, 1109,
  \dodoi{10.1086/521813}

\bibitem[{S.~E. {Woosley} {et~al.}(1986){Woosley}, {Taam}, \&
  {Weaver}}]{woosley1986ApJ...301..601W}
{Woosley}, S.~E., {Taam}, R.~E., \& {Weaver}, T.~A. 1986,
  \bibinfo{title}{{Models for Type I Supernova. I. Detonations in White
  Dwarfs},} \apj, 301, 601, \dodoi{10.1086/163926}

\bibitem[{H. Yamaguchi {et~al.}(2015)Yamaguchi, Badenes, Foster, Bravo,
  Williams, Maeda, Nobukawa, Eriksen, Brickhouse, Petre,
  {et~al.}}]{yamaguchi2015chandrasekhar}
Yamaguchi, H., Badenes, C., Foster, A.~R., {et~al.} 2015, \bibinfo{title}{A
  Chandrasekhar mass progenitor for the Type Ia supernova remnant 3C 397 from
  the enhanced abundances of nickel and manganese,} The Astrophysical Journal
  Letters, 801, L31

\bibitem[{H. {Yamaguchi} {et~al.}(2015){Yamaguchi}, {Badenes}, {Foster},
  {Bravo}, {Williams}, {Maeda}, {Nobukawa}, {Eriksen}, {Brickhouse}, {Petre},
  \& {Koyama}}]{yamaguchietal15}
{Yamaguchi}, H., {Badenes}, C., {Foster}, A.~R., {et~al.} 2015,
  \bibinfo{title}{{A Chandrasekhar Mass Progenitor for the Type Ia Supernova
  Remnant 3C 397 from the Enhanced Abundances of Nickel and Manganese},} \apjl,
  801, L31, \dodoi{10.1088/2041-8205/801/2/L31}

\bibitem[{M. {Zingale} {et~al.}(2009){Zingale}, {Almgren}, {Bell}, {Nonaka}, \&
  {Woosley}}]{zingaleetal09}
{Zingale}, M., {Almgren}, A.~S., {Bell}, J.~B., {Nonaka}, A., \& {Woosley},
  S.~E. 2009, \bibinfo{title}{{Low Mach Number Modeling of Type IA Supernovae.
  IV. White Dwarf Convection},} \apj, 704, 196,
  \dodoi{10.1088/0004-637X/704/1/196}

\bibitem[{M. Zingale {et~al.}(2011)Zingale, Nonaka, Almgren, Bell, Malone, \&
  Woosley}]{zingale2011convective}
Zingale, M., Nonaka, A., Almgren, A., {et~al.} 2011, \bibinfo{title}{The
  convective phase preceding type ia supernovae,} The Astrophysical Journal,
  740, 8

\bibitem[{M. Zingale {et~al.}(2005)Zingale, Woosley, Rendleman, Day, \&
  Bell}]{Zingale2005}
Zingale, M., Woosley, S.~E., Rendleman, C.~A., Day, M.~S., \& Bell, J.~B. 2005,
  \bibinfo{title}{Three‐dimensional Numerical Simulations of
  Rayleigh‐Taylor Unstable Flames in Type Ia Supernovae,} The Astrophysical
  Journal, 632, 1021–1034, \dodoi{10.1086/433164}

\end{thebibliography}
\end{document}